\documentclass[12pt, onecolumn]{article}
\usepackage[dvips]{graphics}
\usepackage[dvips]{color}
\usepackage{amssymb}
\usepackage{amsmath}
\usepackage{amsthm}
\usepackage[dvips]{geometry}
\usepackage{enumitem}

\DeclareMathSymbol{\R}{\mathalpha}{AMSb}{"52}
\newtheorem{thm}{Theorem}
\newtheorem{lemma}{Lemma}

\def\tr{\mathop{\rm tr}\nolimits}%
\def\erfc{\mathop{\rm erfc}\nolimits}%
\def\sgn{\mathop{\rm sgn}\nolimits}%
\def\var{\mathop{\rm Var}\nolimits}%
\def\cov{\mathop{\rm Cov}\nolimits}%
\def\spanop{\mathop{\rm span}\nolimits}%
\def\argmax{\text{max}^{-1}}%

\begin{document}

\title{Feedback Capacity of the First-Order Moving Average Gaussian
Channel}

\author{Young-Han Kim\thanks{This work was supported in part by NSF
Grant CCR-0311633.}\\
Information Systems Laboratory\\
Stanford University}
\maketitle

\begin{abstract}
The feedback capacity of the stationary Gaussian additive noise
channel has been open, except for the case where the noise is white.
Here we find the feedback capacity of the stationary first-order
moving average additive Gaussian noise channel in closed form.
Specifically, the channel is given by $Y_i = X_i + Z_i,$ $i = 1, 2,
\ldots,$ where the input $\{X_i\}$ satisfies a power constraint and
the noise $\{Z_i\}$ is a first-order moving average Gaussian process
defined by $Z_i = \alpha\,U_{i-1} + U_i,\;|\alpha| \le 1,$ with white
Gaussian innovations $U_i,$ $i = 0,1,\ldots.$ 

We show that the feedback capacity of this channel is $-\log x_0,$
where $x_0$ is the unique positive root of the equation $ \rho\,x^2 =
(1-x^2) (1 - |\alpha|x)^2,$ and $\rho\,$ is the ratio of the average
input power per transmission to the variance of the noise innovation
$U_i$.  The optimal coding scheme parallels the simple linear
signalling scheme by Schalkwijk and Kailath for the additive white
Gaussian noise channel --- the transmitter sends a real-valued
information-bearing signal at the beginning of communication and
subsequently refines the receiver's error by processing the feedback
noise signal through a linear stationary first-order autoregressive
filter.  The resulting error probability of the maximum likelihood
decoding decays doubly-exponentially in the duration of the
communication.  This feedback capacity of the first-order moving
average Gaussian channel is very similar in form to the best known
achievable rate for the first-order \emph{autoregressive} Gaussian
noise channel studied by Butman, Wolfowitz, and Tiernan, although the
optimality of the latter is yet to be established.
\end{abstract}

\section{Introduction and Summary}
\label{sec-intro}

Consider the additive Gaussian noise channel with feedback as
depicted in Figure~1.
\begin{figure}[ht]
\begin{center}
\input{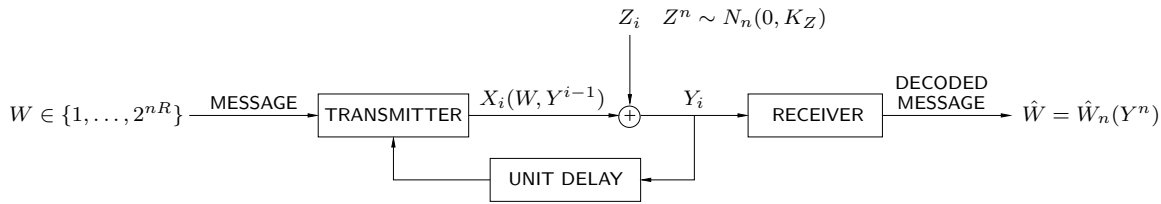}
\caption{Gaussian channel with feedback.}
\end{center}
\end{figure}
The channel $Y_i = X_i + Z_i,$ $i=1,2,\ldots,$ has additive Gaussian
noise $Z_1,Z_2,\ldots,$ where $Z^n = (Z_1,\ldots,Z_n) \sim N_n(0,
K_Z)$.  We wish to communicate a message $W \in \{1,2,\ldots,
2^{nR}\}$ reliably over the channel $Y^n = X^n + Z^n$.  The channel
output is causally fed back to the transmitter.  We specify a
$(2^{nR}, n)$ code with the codewords\footnote{More precisely,
encoding functions $X_i : \{1,\ldots,2^{nR}\} \times \R^{i-1} \to
\R,\,i=1,2,\ldots,n$.}  $(X_1(W),$ $X_2(W,Y_1),$ $\ldots,$
$X_n(W,Y^{n-1}))$ satisfying the expected power constraint
\begin{align}
E \frac{1}{n} \sum_{i=1}^n X_i^2(W,Y^{i-1}) \le P
\label{power const}
\end{align}
and decoding function $\hat{W}_n: \R^n \to
\{1,2,\ldots,2^{nR}\}.$ The probability of error $P_e^{(n)}$ is
defined by
\begin{align*}
P_e^{(n)} := \Pr \{ \hat{W}_n(Y^n) \ne W \}
\end{align*}
where the message $W$ is independent of $Z^n$ and is uniformly
distributed over $\{1,2,\ldots,2^{nR}\}$.  We call the sequence
$\{C_{n,\text{FB}}\}_{n=1}^\infty$ `` an $n$-block feedback capacity
sequence if for every $\epsilon > 0$, there exists a sequence of
$(2^{n(C_{n,\text{FB}}-\epsilon)}, n)$ codes with $P_e^{(n)} \to 0$ as
$n \to \infty$, and for every $\epsilon > 0$ and any sequence of codes
with $2^{n(C_{n,\text{FB}}+ \epsilon)}$ codewords, $P_e^{(n)}$ is
bounded away from zero for all $n$.  We define the feedback capacity
$C_{\text{FB}}$ as
\begin{align*}
C_\text{FB} := \lim_{n\to\infty} C_{n,\text{FB}}
\end{align*}
if the limit exists.  This definition of feedback capacity agrees with
the usual operational definition for the capacity of memoryless
channels without feedback as the supremum of achievable rates~\cite{CT
1991}.

In \cite{CP 1989}, Cover and Pombra 
characterized
the $n$-block feedback capacity $C_{n,\text{FB}}$
as
\begin{align}
\label{fb cap 1}
C_{n,\text{FB}} = \max_{\tr(K_X) \le nP}
\frac{1}{2n} \log \frac{
\det(K_Y)}{\det(K_Z)}.
\end{align}
Here $K_X = K_X(n)$, $K_Y = K_Y(n)$ and $K_Z=K_Z(n)$ respectively
denote the covariance matrices of $X^n,$ $Y^n$ and $Z^n$, and the
maximization is over all $X^n$ of the form $X^n = B Z^n + V^n$ with a
strictly lower-triangular $n \times n$ matrix $B = B(n)$ and
multivariate Gaussian $V^n$ independent of $Z^n$ such that $E
\sum_{i=1}^n X_i^2 = \tr K_X \le nP.$ Equivalently, we can rewrite
(\ref{fb cap 1}) as
\begin{align}
\label{fb cap 2}
C_{n,\text{FB}} = \max_{K_V, B} \frac{1}{2n} \log \frac{
\det((B+I)K_Z(B+I)^T + K_V)}
{\det(K_Z)}
\end{align}
where the maximization is over all nonnegative definite $n \times n$
matrices $K_V = K_V(n)$ and strictly lower triangular $n\times n$
matrices $B = B(n)$ such that $ \tr(BK_ZB^T + K_V) \le nP.  $

When the noise process $\{Z_n\}$ is stationary, the $n$-block
capacity is super-additive in the sense that
\begin{align*}
n\, C_{n,\text{FB}} + m\, C_{m,\text{FB}} \le
(n+m)\, C_{n+m,\text{FB}}, \quad\text{for all } n,m = 1,2,\ldots.
\end{align*}
Consequently, the feedback capacity $C_{\text{FB}}$ is well-defined
(see, for example, P\'olya and Szeg\"o~\cite{PS
1976}) as
\begin{align}
C_{\text{FB}} &= \lim_{n\to\infty}C_{n,\text{FB}}\nonumber\\ &=
\lim_{n\to\infty} \max_{B{(n)},K_V{(n)}} \frac{1}{2n} \log \frac{
\det((B+I)K_Z(B+I)^T + K_V)}{\det(K_Z)}.
\label{fb cap 3} 
\end{align}
To obtain a closed-form expression for the feedback capacity
$C_\text{FB}$, however, we need to go further than (\ref{fb cap 3})
since the above characterization does not give any hint on the
sequence of the optimal $(B(n), K_V(n))_{n=1}^\infty$ achieving
$C_{n,\text{FB}}$ or more importantly, its limiting behavior.

In this paper, we study in detail the case where the additive
Gaussian noise process $\{Z_i\}_{i=1}^\infty$ is a moving average
process of order one (MA(1)).
We define the Gaussian
MA(1) noise process $\{Z_i\}_{i=1}^\infty$ with parameter
$\alpha,\;|\alpha| \le 1,$ as
\begin{align}
\label{ma1 noise}
Z_i = \alpha\,U_{i-1} + U_i
\end{align}
where $\{U_i\}_{i=0}^\infty$ is a white Gaussian innovation process.
Without loss of generality, we will assume that $U_i,\:i=0,1,\ldots$,
has unit variance.  There are alternative ways of defining Gaussian
MA(1) processes, which we will review in Section~\ref{sec-noise}.

Note that the condition $|\alpha| \le 1$ is not restrictive.  When
$|\alpha| > 1$, it can be readily verified that the process $\{Z_i\}$
has the same distribution as the process $\{\tilde{Z}_i\}$ defined by
\begin{align*}
\tilde{Z}_i = \alpha (\beta\,U_{i-1} + U_i)
\end{align*}
where the moving average parameter $\beta$ is given by $\beta =
1/\alpha,$ thus giving $|\beta| < 1.$ 

We state the main theorem, the proof of which will be given in
Section~\ref{sec-pf}.
\begin{thm}
For the additive Gaussian MA(1) noise channel $Y_i = X_i + Z_i, \; i =
1,2,\ldots,$ with the Gaussian MA(1) noise process $\{Z_i\}$ defined
in (\ref{ma1 noise}), the feedback capacity $C_\text{FB}$ under the
power constraint $\sum_{i=1}^n E X_i^2 \le n P$ is given by
\begin{align*}
C_\text{FB} = - \log x_0,
\end{align*}
where $x_0$ is the unique positive root of the fourth-order polynomial
\begin{align}
\label{ma1 cap}
P\,x^2 = (1-x^2) (1 - |\alpha| x)^2.
\end{align}
\end{thm}

As will be shown later in Sections 3 and 4, the feedback capacity
$C_\text{FB}$ is achieved by an asymptotically stationary ergodic
input process $\{X_i\}$ satisfying $E X_i^2 = P$ for all $i$.  Thus by
ergodic theorem, the feedback capacity does not diminish under a more
restrictive power constraint
\begin{align*}
\frac{1}{n} \sum_{i=1}^n X_i^2(W,Y^{i-1}) \le P.
\end{align*}
(See also the arguments given in~\cite[Section VIII]{CP 1989}
based on the stationarity of the noise process.) 

The literature on Gaussian feedback channels is vast.  We first
mention some prior work closely related to our main discussion.  In
earlier work, Schalkwijk and Kailath~\cite{SK 1966, Schalkwijk 1966}
(see also the discussion by Wolfowitz~\cite{Wolfowitz 1968})
considered the feedback over the additive white Gaussian noise
channel, and proposed a simple linear signalling scheme that achieves
the feedback capacity.  The coding scheme by Schalkwijk and Kailath
can be summarized as follows: Let $\theta$ be one of $2^{nR}$ equally
spaced real numbers on some interval, say, $[0,1]$.  At time $k$, the
receiver forms the maximum likelihood estimate
$\hat\theta_k(Y_1,\ldots,Y_k)$ of $\theta$.  Using the feedback
information, at time $k+1$, we send $X_{k+1} = \gamma_k (\theta -
\hat\theta_k)$, where $\gamma_k$ is a scaling factor properly chosen
to meet the power constraint.  After $n$ transmissions, the receiver
finds the value of $\theta$ among $2^{nR}$ alternatives that is
closest to $\hat\theta_n$.  This simple signalling scheme, without any
coding, achieves the feedback capacity.  As is shown by
Shannon~\cite{Shannon 1956}, feedback does not increase the capacity
of memoryless channels.  (See also Kadota et al.~\cite{Kadota 1971a,
Kadota 1971b} for continuous cases.)  The benefit of feedback,
however, does not consist of the simplicity of coding only.  The
probability of decoding error of the Schalkwijk-Kailath scheme decays
doubly exponentially in the duration of communication, compared to the
exponential decay for the nonfeedback scenario.  In fact, there exists
a feedback coding scheme such that the probability of decoding error
decreases more rapidly than the exponential of any order~\cite{Pinsker
1968, Kramer 1969, Zigangirov 1970}.  Later Schalkwijk extended his
work to the center-of-gravity information feedback for higher
dimensional signal spaces~\cite{Schalkwijk 1968}.

Butman~\cite{Butman 1969} generalized the linear coding scheme of
Schalkwijk and Kailath for white noise processes to autoregressive
(AR) noise processes.  For first-order autoregressive (AR(1))
processes $\{Z_i\}_{i=1}^\infty$ with regression parameter $\alpha,$
$|\alpha| < 1,$ defined by
\begin{align*}
Z_i = \alpha Z_{i-1} + U_i
\end{align*}
he obtained a lower bound on the feedback capacity as $- \log x_0$,
where $x_0$ is the unique positive root of the fourth-order polynomial
\begin{align}
\label{ar1 cap}
P\,x^2 = \frac{(1-x^2)}{(1 + |\alpha| x)^2}.
\end{align}
This rate has been shown to be optimal among a certain class of linear
feedback schemes by Wolfowitz~\cite{Wolfowitz 1975} and
Tiernan~\cite{Tiernan 1976} and is strongly believed to be the
capacity of the AR(1) feedback capacity.  Tiernan and
Schalkwijk~\cite{TS 1974} found an upper bound of the AR(1) feedback
capacity, which meets Butman's lower bound for very low and very high
signal-to-noise ratio.  Butman~\cite{Butman 1976} also obtained
capacity upper and lower bounds for AR processes with higher order.

For the case of moving average (MA) noise processes, there are far
fewer results in the literature, although MA processes are usually
more tractable than AR processes of the same order.
Ozarow~\cite{Ozarow 1990a, Ozarow 1990b} gave upper and lower bounds
of the feedback capacity for AR(1) and MA(1) channels and showed that
feedback strictly increases the capacity.  Substantial progress was
made by Ordentlich~\cite{Ordentlich 1995a}; he observed that $K_V$ in
(\ref{fb cap 2}) is at most of rank $k$ for a MA noise process with
order $k$.  He also showed that the optimal $(K_V, B)$ necessarily has
the property that the current input signal $X_k$ is orthogonal to the
past outputs $(Y_1,\ldots,Y_{k-1})$.  For the special case of MA(1)
processes, this development, combined with the arguments given
in~\cite{Wolfowitz 1975}, suggests that a linear signalling scheme
similar to the Schalkwijk-Kailath scheme be optimal, which is proved
by our Theorem~1.

A recent report by Yang, Kav\v{c}i\'{c}, and Tatikonda~\cite{YKT 2004}
(see also Yang's thesis~\cite{Yang 2004}) studies the feedback
capacity of the general ARMA($k$) case using the state-space model and
offers a conjecture on the feedback capacity as a solution to an
optimization problem that does not depend on the horizon $n$.  For the
special case $k = 1$ with the noise process $\{Z_i\}_{i=1}^\infty$
defined by
\begin{align*}
Z_i = \beta\,Z_{i-1} + \alpha\,U_{i-1} + U_i,\qquad |\alpha|,|\beta| < 1
\end{align*}
they conjecture that the Schalkwijk-Kailath-Butman scheme is
optimal.  The corresponding achievable rate can be written in a closed
form as $- \log x_0$, where $x_0$ is the unique positive root of the
fourth-order polynomial
\begin{align*}
P\,x^2 = \frac{(1-x^2)(1-\sigma \alpha x)^2}{(1 + \sigma\beta x)^2}
\end{align*}
and
\[
\sigma = \left\{
\begin{array}{ll}
1,&\quad \alpha+\beta \ge 0,\\
-1,&\quad \alpha+\beta < 0.
\end{array}\right.
\]
By taking $\beta = 0$ or $\alpha = 0$, we can easily recover
\eqref{ma1 cap} and \eqref{ar1 cap}, respectively.  Thus, in the
special case $\beta = 0$, our Theorem~1 confirms the
Yang-Kav\v{c}i\'{c}-Tatikonda conjecture.

To conclude this section, we review, in a rather incomplete manner,
previous work on the Gaussian feedback channel in addition to
aforementioned results, and then point out where the current work lies
in the literature.  The standard literature on the Gaussian feedback
channel and associated simple feedback coding schemes traces back to a
1956 paper by Elias~\cite{Elias 1956} and its sequels~\cite{Elias
1961, Elias 1967}.  Turin~\cite{Turin 1965, Turin 1966, Turin 1968},
Horstein~\cite{Horstein 1966}, Khas'minskii~\cite{Khas'minskii 1967},
and Ferguson~\cite{Ferguson 1968} studied a sequential binary
signalling scheme over the Gaussian feedback channel with
symbol-by-symbol decoding that achieves the feedback capacity with an
error exponent better than the nonfeedback case.  As mentioned above,
Schalkwijk and Kailath~\cite{SK 1966, Schalkwijk 1966, Schalkwijk
1968} made a major breakthrough by showing that a simple linear
feedback coding scheme achieves the feedback capacity with doubly
exponentially decreasing probability of decoding error.  This
fascinating result has been extended in many directions.
Omura~\cite{Omura 1968} reformulated the feedback communication
problem as a stochastic-control problem and applied this approach to
multiplicative and additive noise channels with noiseless feedback and
to additive noise channels with noisy feedback.  Pinsker~\cite{Pinsker
1968}, Kramer~\cite{Kramer 1969}, and Zigangirov~\cite{Zigangirov
1970} studied feedback coding schemes under which the probability of
decoding error decays as the exponential of arbitrary high order.
Wyner~\cite{Wyner 1968} and Kramer~\cite{Kramer 1969} studied the
performance of the Schalkwijk-Kailath scheme under a peak power
constraint and reported the singly exponential behavior of the
probability of decoding error under a peak power constraint.  The
actual error exponent of the Gaussian feedback channel under the peak
power constraint was later obtained by Schalkwijk and Barron~\cite{SB
1971}.  Kashyap~\cite{Kashyap 1968}, Lavenberg~\cite{Lavenberg 1969,
Lavenberg 1971} and Kramer~\cite{Kramer 1969} looked at the case of
noisy or intermittent feedback.

The more natural question of transmitting a Gaussian source over a
Gaussian feedback channel was studied by Kailath~\cite{Kailath 1967},
Cruise~\cite{Cruise 1967}, Schalkwijk and Bluestein~\cite{SB 1967},
Ovseevich~\cite{Ovseevich 1970}, and Ihara~\cite{Ihara 1973}.  There
are also many notable extensions of the Schalkwijk-Kailath scheme in
the area of multiple user information theory.  Using the
Schalkwijk-Kailath scheme, Ozarow and Leung-Yan-Cheong~\cite{OL 1984}
showed that feedback increases the capacity region of
\emph{stochastically} degraded broadcast channels, which is rather
surprising since feedback does \emph{not} increase the capacity region
of \emph{physically} degraded broadcast channels, as shown by El
Gamal~\cite{El Gamal 1978}.  Ozarow~\cite{Ozarow 1984} also
established the feedback capacity region of two-user white Gaussian
multiple access channel through a very innovative application of the
Schalkwijk-Kailath coding scheme.  The extension to a larger number of
users was attempted by Kramer~\cite{Kramer 2002}, where he also showed
that feedback increases the capacity region of strong interference
channels.

Following these results on the white Gaussian noise channel on hand,
the next focus was on the feedback capacity of the colored Gaussian
noise channel.  Butman~\cite{Butman 1969, Butman 1976} extended the
Schalkwijk-Kailath coding scheme to autoregressive noise channels.
Subsequently, Tiernan and Schalkwijk~\cite{TS 1974, Tiernan 1976},
Wolfowitz~\cite{Wolfowitz 1975}, Ozarow~\cite{Ozarow 1990a, Ozarow
1990b}, Dembo~\cite{Dembo 1989}, and Yang et al.~\cite{YKT 2004}
studied the feedback capacity of finite-order ARMA additive Gaussian
noise channels and obtained many interesting upper and lower bounds.
Using an asymptotic equipartition theorem for nonstationary nonergodic
Gaussian noise processes, Cover and Pombra~\cite{CP 1989} obtained the
$n$-block capacity (\ref{fb cap 2}) for the arbitrary colored Gaussian
channel with or without feedback.  (We can take $B = 0$ in \eqref{fb
cap 2} for the nonfeedback case.)  Using matrix inequalities, they
also showed that feedback does not increase the capacity much; namely,
feedback increases the capacity at most twice (a result obtained by
Pinsker~\cite{Pinsker 1969} and Ebert~\cite{Ebert 1970}), and feedback
increases the capacity at most by half a bit.

The extensions and refinements of the result by Cover and Pombra
abound.  Dembo~\cite{Dembo 1989} showed that the feedback does not
increase the capacity at very low signal-to-noise ratio or very high
signal-to-noise ratio.  As mentioned above,
Ordentlich~\cite{Ordentlich 1995a} examined the properties of the
optimal solution $(K_V, B)$ in (\ref{fb cap 2}) and found the rank
condition of $K_V$ for finite-order MA noise processes.  Chen and
Yanagi~\cite{Yanagi 1994, CY 1999, CY 2000} studied Cover's
conjecture~\cite{Cover 1987} that the feedback capacity is at most as
large as the nonfeedback capacity with twice the power, and made
several refinements on the upper bounds by Cover and Pombra.
Thomas~\cite{Thomas 1987}, Pombra and Cover~\cite{PC 1994}, and
Ordentlich~\cite{Ordentlich 1996} extended the factor-of-two bound
result to the colored Gaussian multiple access channels with feedback.
Recently Yang, Kav\v{c}i\'{c}, and Tatikonda~\cite{YKT 2004} revived
the control-theoretic approach (\text{cf.} \cite{Omura 1968}) to the
stationary ARMA($k$) Gaussian feedback capacity problem.  Although
one-sentence summary would not do justice to their contribution, Yang
et \text{al.} reformulated the feedback capacity problem as a
stochastic control problem and used dynamic programming for the
numerical computation of the $n$-block feedback capacity.  In a series
of papers~\cite{Ihara 1980, Ihara 1990, Ihara 1994}, Ihara obtained
coding theorems for continuous-time Gaussian channels with feedback
and showed that the factor-of-two bound on the feedback capacity is
tight by considering cleverly constructed nonstationary channels both
in discrete time~\cite{Ihara 1988} and continuous time~\cite{Ihara
1990}.  (See also~\cite[Examples 5.7.2 and 6.8.1]{Ihara 1993}.)  In
fact, besides the white Gaussian noise channel, Ihara's example is the
only nontrivial channel with known closed-form feedback capacity.

Hence Theorem~1 provides the first feedback capacity result on
stationary colored Gaussian channels.  Moreover, as will be discussed
in Section~4, a simple linear signalling scheme similar to the
Schalkwijk-Kailath scheme achieves the feedback capacity.  This result
links the Cover-Pombra formulation of the feedback capacity with the
Schalkwijk-Kailath scheme and its generalizations to stationary
colored channels, and provides new hope for the optimality of the
achievable rate for the AR(1) channel obtained by Butman~\cite{Butman
1969}.

\section{First-Order Moving Average Gaussian Processes}
\label{sec-noise}
In this section, we digress a little to review a few characteristics
of first-order moving average Gaussian processes.  First, we give
three alternative characterizations of Gaussian MA(1) processes.
As defined in the previous section, the Gaussian
MA(1) noise process $\{Z_i\}_{i=1}^\infty$ with parameter $\alpha$
can be characterized as
\begin{align}
\label{ma1 noise2}
Z_i = \alpha\,U_{i-1} + U_i,
\end{align}
where the innovations $U_0, U_1,\ldots$ are
\text{i.i.d.} $\sim N(0,1).$

We reinterpret the above definition in (\ref{ma1 noise2}) by regarding
the noise process $\{Z_i\}$ as the output of the linear time-invariant
filter with transfer function
\begin{align}
\label{ma1 filter}
H(z) = 1 + \alpha z^{-1},
\end{align}
which is driven by the white innovation process $\{U_i\}$.  Thus we
alternatively characterize the Gaussian MA(1) noise process $\{Z_i\}$
with parameter $\alpha$ and unit innovation through its power spectral
density $S_Z(\omega)$ given by
\begin{align}
\label{ma1 spectrum}
S_Z(\omega)
= |1 + \alpha e^{-j\omega}|^2 = 1 + \alpha^2 + 2\alpha \cos \omega.
\end{align}

We can further identify the power spectral density $S_Z(\omega)$ with
the infinite Toeplitz covariance matrix of a Gaussian process.
Thus, we can define $\{Z_i\}$ as $(Z_1,\ldots,Z_n) \sim N_n(0,K_Z)$
for each finite horizon $n$ where $K_Z$ is tri-diagonal with
\begin{gather*}
K_Z = 
\left[
\begin{matrix}
1+\alpha^2 & \alpha & 0 & \cdots & 0\\
\alpha & 1+\alpha^2 & \alpha & \ddots & \vdots\\
0 & \alpha & 1+\alpha^2 & \ddots & 0 \\
\vdots & \ddots & \ddots & \ddots & \alpha\\
0 & \cdots  & 0 & \alpha & 1+\alpha^2
\end{matrix}
\right],
\intertext{or equivalently,}
[K_Z]_{i,j}
=
\left\{
\begin{array}{ll}
1 + \alpha^2,&\quad |i - j| = 0,\\
\alpha,&\quad |i - j| = 1,\\
0, &\quad |i-j| \ge 2.
\end{array}
\right.
\end{gather*}
Note that this covariance matrix $K_Z$ is consistent with our initial
definition of the MA(1) process given in (\ref{ma1 noise2}).  Thus all
three definitions of the MA(1) process given above are equivalent.  As
we will see in the next section, the special structure of the MA(1)
process, especially the tri-diagonality of the covariance matrix, makes
the maximization in (\ref{fb cap 2}) easier than the generic case.

We will need the entropy rate of the MA(1) Gaussian process later in
our discussion.  As shown by Kolmogorov (see~\cite[Section 11.6]{CT
1991}), the entropy rate of a stationary Gaussian process with power
spectral density $S(\omega)$ can be expressed as
\begin{align*}
\frac{1}{4\pi} \int_{-\pi}^{\pi} \log \left(2\pi e S(\omega)\right) d\omega.
\end{align*}
We can calculate the above integral with the power spectral density
$S_Z(\omega)$ in (\ref{ma1 spectrum}) by Jensen's\footnote{The same
J.~L.~W.~V.~Jensen famous for his inequality on convex functions.}
formula~\cite[Theorem 15.18]{Rudin}
\begin{equation}
\label{jensen}
\frac{1}{2\pi} \int_{-\pi}^{\pi}
\log | e^{j\omega} - \alpha | \,d\omega
= \left\{
\begin{array}{ll}
0 ,&\quad |\alpha| \le 1,\\[.25em]
\log |\alpha|,&\quad |\alpha| > 1,
\end{array}
\right.
\end{equation}
and obtain the entropy rate of the
MA(1) Gaussian process (\ref{ma1 noise2}) as
\begin{align}
\nonumber
\frac{1}{4\pi} \int_{-\pi}^{\pi} \log \left(2\pi e S_Z(\omega)\right)
d\omega 
&= \frac{1}{4\pi} \int_{-\pi}^{\pi} \log \left(2\pi e | 1 + \alpha e^{-j\omega} | ^2 \right)
d\omega\\[.5em]
&= \left\{
\begin{array}{ll}
\frac{1}{2} \log (2\pi e),&\quad |\alpha| \le 1,\\[1em]
\frac{1}{2} \log (2\pi e \alpha^2),&\quad |\alpha| > 1.
\end{array}
\right.
\label{ent rate}
\end{align}
(One can alternatively deal with the
determinant of $K_Z(n)$ directly by a simple recursion.  For example,
we can show that $\det K_Z(n) = n+1$ for $|\alpha| = 1$.)  For a more
general discussion of the entropy rate of stationary Gaussian
processes, refer to \cite[Chapter 2]{Ihara 1993}.

We finish our digression by noting a certain reciprocal relationship
between the Gaussian MA(1) process with parameter $\alpha$ and the
Gaussian AR(1) process with parameter $-\alpha$.  We can define the
Gaussian AR(1) process $\{Z_i\}_{i=1}^\infty$ with parameter $-\alpha,$
$|\alpha| < 1,$ as
\begin{align*}
Z_i = - \alpha Z_{i-1} + U_i,
\end{align*}
where the innovations $U_1,U_2, \ldots$ are \text{i.i.d.}  $\sim
N(0,1)$ and $Z_0 \sim N(0, 1/(1-\alpha^2))$ is independent of
$\{U_i\}_{i=1}^\infty$.  Equivalently, we can define the above process as
the output of the linear time-invariant filter with transfer function
\begin{align*}
G(z) = \frac{1}{1 + \alpha z^{-1}} = \frac{1}{H(z)},
\end{align*}
where $H(z)$ is the transfer function (\ref{ma1 filter}) of the MA(1)
process with parameter $\alpha$.  
This reciprocity 
is indeed reflected in the striking similarity between the
fourth-order polynomial (\ref{ma1 cap}) for the capacity of the
Gaussian MA(1) noise channel and the fourth-order polynomial (\ref{ar1
cap}) for the best known achievable rate of the Gaussian AR(1) noise
channel.

\section{Proof of Theorem 1}
\label{sec-pf}

We will first transform the optimization problem 
\begin{align*}
C_{n,\text{FB}} = \max_{K_V, B} \frac{1}{2n} \log \frac{
\det((B+I)K_Z(B+I)^T + K_V)}
{\det(K_Z)}
\tag{\ref{fb cap 2}}
\end{align*}
to a series of (asymptotically) equivalent forms.  Then we solve the
problem by imposing individual power constraints $(P_1,\ldots, P_n)$
on each input signal.  Subsequently we optimize over $(P_1,\ldots,
P_n)$ under the average power constraint
\begin{align*}
P_1 + \cdots + P_n \le n P.
\end{align*}
Then using Lemma~\ref{lemma asymp}, we will prove that the uniform
power allocation $P_1 = \cdots = P_n = P$ is asymptotically optimal.
This leads to a closed-form solution given in Theorem~1.

\vskip 1em
{\it Step 1. Transformations into equivalent optimization problems.}

Recall that we
wish to solve the optimization problem:
\begin{align}
\label{opt 1}
\text{maximize}\quad \log
\det((B+I)K_Z(B+I)^T + K_V)
\end{align}
over all nonnegative definite $K_V$ and strictly lower triangular $B$
satisfying $\tr(BK_ZB^T + K_V) \le nP.$ We approximate the covariance
matrix $K_Z$ of the given MA(1) noise process with parameter $\alpha$
by another covariance matrix $K'_{{Z}}$.  Define $K'_{{Z}} =
{H}_Z {H}_Z^T$ where the lower-triangular Toeplitz matrix ${H}_Z$ is
given by
\begin{align*}
{H}_Z = \left[
\begin{array}{ccccc}
1& 0 & 0& \cdots& 0\\[3pt]
\alpha& 1 & 0 &\cdots&0\\
0 & \alpha & 1 & \ddots& \vdots\\[3pt]
\vdots & \ddots& \ddots& \ddots& 0 \\[3pt]
0 &\cdots & 0 & \alpha & 1
\end{array}
\right].
\end{align*}
This matrix ${K}'_{{Z}}$ is a covariance matrix of the 
Gaussian process $\{\tilde{Z}_i\}_{i=0}^\infty$ defined by
\begin{align*}
\tilde{Z}_1 &= U_1,\\
\tilde{Z}_i &= U_i + \alpha\,U_{i-1},\quad i = 2, 3,\ldots,
\end{align*}
where $\{U_i\}_{i=1}^\infty$ is the white Gaussian process with unit
variance.  It is easy to check that $K_Z \succeq K'_{{Z}}$ (i.e., $K_Z
- K'_Z$ is nonnegative definite) and that the difference between $K_Z$
and $K'_{{Z}}$ is given by
\begin{align*}
[\,K_Z - K'_{{Z}}]_{i,j} =
\left\{
\begin{array}{ll}
\alpha^2,&\quad i=j=1,\\
0,&\quad\text{otherwise}.
\end{array}\right.
\end{align*}
It is intuitively clear that there is no asymptotic difference in
capacity between the channel with the original noise covariance $K_Z$
and the channel with noise covariance $K'_{{Z}}$.  We will prove this
claim more rigorously in the Appendix.  Throughout we will assume that
the noise covariance matrix of the given channel is $K'_{{Z}}$, which
is equivalent to the statement that the time-zero noise innovation
$U_0$ is revealed to both the transmitter and the receiver.

Now by identifying $K_V = {F}_V {F}_V^T$
for some lower-triangular ${F}_V$ and identifying ${F}_Z = B {H}_Z$
for some strictly lower-triangular ${F}_Z$, we transform the
optimization problem (\ref{opt 1}) into
\begin{align}
\label{opt 2}
\begin{array}{ll}
\text{maximize}&\quad \log\det({F}_V {F}_V^T +
({F}_Z + {H}_Z) ({F}_Z + {H}_Z)^T)\\[6pt]
\text{subject to}&\quad \tr({F}_V {F}_V^T + {F}_Z {F}_Z^T) \le nP
\end{array}
\end{align}
with new variables $({F}_V,{F}_Z)$.

We shall use $2n$-dimensional row vectors $f_i$ and $h_i$, $i =
1,\ldots,n$, to denote the $i$-th row of $F := [F_V ~ F_Z]$ and $H :=
[\,0_{n\times n} ~ H_Z]$, respectively.  There is an obvious
identification between the time-$i$ input signal $X_i$ and the vector
$f_i$, $i=1,\ldots,n,$ for we can regard $f_i$ as a point in the
Hilbert space with the innovations of $V^n$ and $Z^n$ as a basis.  We
can similarly identify $Z_i$ with $h_i$ and identify $Y_i$ with $f_i +
h_i$.  We also introduce new variables $(P_1,\ldots,P_n)$ representing
the power constraint for each input $f_i$.  Now the optimization
problem in (\ref{opt 2}) becomes the following equivalent form:
\begin{align}
\label{opt 4}
\begin{array}{ll}
\text{maximize}&\quad \log \det((F + H)(F + H)^T)\\[6pt]
\text{subject to}&\quad \|f_i\|^2 \le P_i, \quad i=1,\ldots,n,\\[6pt]
&\quad \sum_{i=1}^n P_i \le nP.
\end{array}
\end{align}
Here $\|\cdot\|$ denotes the Euclidean norm of a $2n$-dimensional
vector.  Note that the variables $f_1,\ldots,f_n$ should satisfy $f_i
\in \mathcal{V}_i,\; i = 1,\ldots, n,$ where
\begin{align*}
\mathcal{V}_i := \{(v_1,\ldots,v_{2n}) \in \R^{2n}:
v_{i+1} = \cdots = v_{n} = 0 = v_{n + i} = \cdots = v_{2n}\}.
\end{align*}

\vskip 1em
{\it Step 2. Optimization under the individual power constraint for
each signal.}

We solve the optimization problem (\ref{opt 4}) in $(f_1,\ldots,f_n)$
after fixing $(P_1,\ldots, P_n)$.  This step is mostly algebraic, but
we can easily give a geometric interpretation.  We need some notation
first.

We define an $n$-by-$2n$ matrix
\begin{align*}
S = 
\left[
\begin{array}{c}
s_1\\
\vdots\\
s_n
\end{array}
\right]
:=
\left[
\begin{array}{c}
f_1 + h_1\\
\vdots\\
f_n + h_n
\end{array}
\right]
= F + H,
\end{align*}
and we define the $n$-by-$2n$ matrix $E$ by
\begin{align*}
E = 
\left[
\begin{array}{c}
e_1\\
\vdots\\
e_n
\end{array}
\right]
:=
[\,0_{n\times n}~I\,],
\end{align*}
where $I$ is identity.  We also define an $n$-by-$2n$ matrix
\begin{align*}
G = 
\left[
\begin{array}{c}
g_1\\
\vdots\\
g_n
\end{array}
\right]
:=
\left[
\begin{array}{c}
h_1 - e_1\\
\vdots\\
h_n - e_n
\end{array}
\right]
= H - E.
\end{align*}
We can interpret the row vector $e_i$ as the noise innovation $U_i$ and 
the row vector $g_i$ as $Z_i - U_i$.

We will use the notation $F_k$ to denote the $k$-by-$2n$ submatrix of $F$
which consists of the first $k$ rows of $F$, that is,
\begin{align*}
F_k = 
\left[
\begin{array}{c}
f_1\\
\vdots\\
f_k
\end{array}
\right].
\end{align*}
We will use the similar notation for the $k$-by-$2n$ submatrices of
$G, H, E,$ and $S$.

We now introduce a sequence of $2n$-by-$2n$ matrices
$\{\Pi_{k}\}_{k=1}^{n-1}$ as
\begin{align*}
\Pi_{k} = I - S_k^T (S_k S_k^T)^{-1} S_k.
\end{align*}
Observe that $S_k$ is of full rank and thus that $(S_k S_k^T)^{-1}$
always exists.  We can view $\Pi_{k}$ as a map of a $2n$-dimensional
row vector (acting from the right) to its component orthogonal to the
subspace spanned by the rows $s_1,\ldots,s_k$ of $S_k$.  (Or $\Pi_k$
maps a generic random variable $A$ to $A - E(A|Y^{k})$.)  It is easy
to verify that $\Pi_k = \Pi_k^T = \Pi_k \Pi_k \text{ and } \Pi_k S_k^T
= 0.$

Finally we define the intermediate objective functions of the
maximization (\ref{opt 4}) as
\begin{gather*}
J_k(P_1,\ldots,P_k) := \max_{\substack{
f_1,\ldots,f_k\\
\|f_i\|^2 \le P_i}}
\log \det (S_k S_k^T),\quad k = 1,\ldots,n,\\
\intertext{so that}
C_{n,\text{FB}} = \max_{P_i:\;\sum P_i \le nP} \frac{1}{2n} J_n (P_1,\ldots, P_n).
\end{gather*}

We will show that if $(f_1^*,\ldots,f_{k-1}^*)$ maximizes
$J_{k-1}(P_1,\ldots, P_{k-1})$, then $(f_1^*, \ldots, f_{k-1}^*,
f_{k}^*)$ maximizes $J_{k}(P_1,\ldots, P_{k})$ for some $f_k^*$
satisfying $f_{k}^* = f_{k}^* \,\Pi_{k-1}.$ Thus the maximization for
$J_n$ can be solved in a greedy fashion by sequentially maximizing
$J_1, J_2, \ldots, J_n$ through $f_1^*, f_2^*, \ldots, f_n^*$.
Furthermore, we will obtain the recursive relationship
\begin{align}
J_0 &:= 0, \label{recursion1}\\
J_1 &= \log (1 + P_1),\label{recursion2}\\
J_{k+1} &= J_k  + 
\log \left(1 + \left(\sqrt{P_{k+1}} + 
|\alpha| \sqrt{1 - \frac{ 1}{e^{J_k - J_{k-1}}}} \right)^2\right),
\quad k = 1, 2, \ldots.\label{recursion3}
\end{align}

We need the following result to proceed to the actual maximization.

\begin{lemma}
Suppose $P \ge 0$ and $1 \le k \le n-1$.  Suppose $S_k$ and $\Pi_k$
defined as above.  Let $\mathcal{V}$ be an arbitrary subspace of\,
$\R^{2n}$ such that $\mathcal{V}$ is not contained in the span of
$s_1,\ldots,s_k$.  Then, for any $w \in \mathcal{V}$,
\begin{align*}
\max_{v \in \mathcal{V}: \|v\|^2 \le P} (v + w) \,\Pi_k (v + w)^T
= \big(\sqrt{P} + \|w \,\Pi_k \|\big)^2.
\end{align*}
Furthermore, if $w \,\Pi_k \neq 0$,
the maximum is attained by
\begin{align}
\label{opt v}
v^* = \sqrt P \frac{w\,\Pi_k}{ \| w\,\Pi_k \| }
\end{align}
\end{lemma}

\begin{proof}
When $w \,\Pi_k = 0$, that is, $w \in \spanop\{s_1,\ldots,s_k\}$, the
maximum of $(v+w)\,\Pi_k(v+w)^T = v\,\Pi_k v^T$ is attained by any
vector $v,$ $\|v\|^2 = P,$ orthogonal to $\spanop\{s_1, \ldots,
s_k\}$, and we trivially have
\begin{align*}
\max_{v \in \mathcal{V}: \|v\|^2 \le P} v \,\Pi_k v^T
= P.
\end{align*}

When $w \,\Pi_k \ne 0$, we have 
\begin{align*}
(v + w) \,\Pi_k (v + w)^T &= \|(v + w \,\Pi_k) \,\Pi_k \|^2 \\
&\le \|v + w \,\Pi_k \|^2 \\
&\le \big(\sqrt{P} + \|w \,\Pi_k \|\big)^2,
\end{align*}
where the first inequality follows from the fact that $I - \Pi_k$ is nonnegative definite.
It is easy to check that we have equality if $v$ is given by (\ref{opt v}).
\end{proof}

\begin{figure}[t]
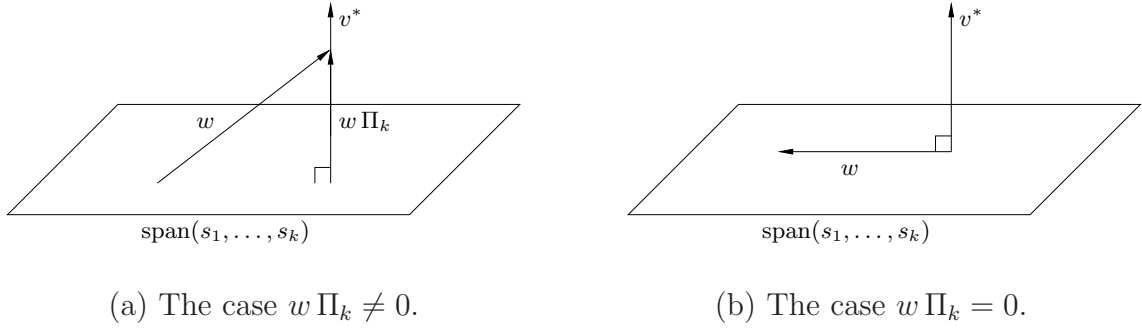

\begin{center}
\begin{tabular}{c@{\hspace{30pt}}c}
\input{proj1.pstex_t}&\hspace{10pt}\input{proj2.pstex_t}\\[12pt]
(a) The case $w\,\Pi_k \ne 0$.&
(b) The case $w\,\Pi_k = 0$.
\end{tabular}
\caption{Geometric interpretation of Lemma 1.}
\end{center}
\end{figure}

We observe that, for $k = 2,\ldots, n$,
\begin{align}
\det (S_k S_k^T) &= \det
\left(\left[
\begin{array}{c}
S_{k-1}\notag\\
s_k
\end{array}
\right] \left[
\begin{array}{c}
S_{k-1}\notag\\
s_k
\end{array}
\right]^T \right)\notag\\
&= \det
\left[
\begin{array}{cc}
S_{k-1}S_{k-1}^T & S_{k-1} s_k^T\notag\\
s_k S_{k-1}^T &s_k s_k^T
\end{array}
\right]\notag\\
&= \det (S_{k-1} S_{k-1}^T) \cdot
s_k (I - S_{k-1}^T (S_{k-1} S_{k-1}^T)^{-1} S_{k-1}) s_k^T \notag\\
&= \det (S_{k-1} S_{k-1}^T) \cdot s_k \,\Pi_{k-1} s_k^T \notag\\
&= \det (S_{k-1} S_{k-1}^T) \cdot (f_k + g_k + e_k) \,\Pi_{k-1} (f_k + g_k + e_k)^T \notag\\
\label{eq orth} &= \det (S_{k-1} S_{k-1}^T) \cdot \left[ 1 + (f_k + g_k) \,\Pi_{k-1} (f_k + g_k)^T \right]
\end{align}
where (\ref{eq orth}) follows since $e_k \,\Pi_{k-1} = e_k,$ $e_k
e_k^T = 1,$ and $e_k g_k^T = e_k f_k^T = 0.$ Now fix $f_1, \ldots,
f_{k-1}$.  Since $\mathcal{V}_k$ is not contained in
$\spanop\{s_1,\ldots,s_{k-1}\}$ and $g_k \in \mathcal{V}_k$, we have
from the above lemma and (\ref{eq orth}) that
\begin{align}
\label{max det}
\max_{f_k: \|f_k\|^2 \le P_k} \det (S_k S_k^T) = \det (S_{k-1}S_{k-1}^T) \cdot
\left(1 + \left(\sqrt{P_k} + \| g_k \,\Pi_{k-1} \|\right)^2 \right).
\end{align}
If $\alpha \ne 0$, the maximum of is attained by
\begin{align}
\label{maximizer}
f_k^* = \sqrt{P_k} \frac{g_k \Pi_{k-1}}{\|g_k \,\Pi_{k-1} \|}.
\end{align}
In the special case $\alpha = 0$, that is, when the noise is white,
we trivially have
\begin{align*}
\max_{f_k: \|f_k\|^2 \le P_k} \det (S_k S_k^T) = \det (S_{k-1}S_{k-1}^T) \cdot
(1 + {P_k}),
\end{align*}
which immediately implies that 
$
J_k = J_{k-1} + \log (1 + P_k) = \sum_{i=1}^k \log (1+P_i),
$
which, in turn, combined with the concavity of the logarithm, implies that
\begin{align*}
C_{n,\text{FB}} = C_{\text{FB}} = \frac{1}{2} \log (1 + P).
\end{align*}

We continue our discussion throughout this step under the assumption
$\alpha \ne 0$.  Until this point we have not used the special
structure of the MA(1) noise process.  Now we rely heavily on it.
We trivially have
\begin{equation}
\label{j1}
J_1 = \max_{f_1} \log (s_1 s_1^T) = \log (1 + P_1),
\end{equation}
Following (\ref{max det}), we have, for $k = 2,\ldots, n,$
\begin{align}
\label{two max}
J_k &= \max_{f_1,\ldots,f_{k-1}} \left[ \log \det (S_{k-1} S_{k-1}^T) + 
\log \left(1 + \left(\sqrt{P_k} + \| g_k \,\Pi_{k-1} \|\right)^2 \right) \right].
\end{align}
We wish to show that both terms in (\ref{two max}) are individually
maximized by the same optimizer 
\begin{align}
\nonumber
(f_1^*,\ldots, f_{k-1}^*) &= \argmax \left(\det (S_{k-1}
S_{k-1}^T)\right)\\
&= \argmax \| g_k \,\Pi_{k-1} \|
\label{argmax}
\end{align}
for $k = 2,\ldots,n$.  Once we establish \eqref{argmax}, the desired
recursion formula \eqref{recursion3} for $J_k$ follows immediately
from the definition of $J_k$ and \eqref{two max}.

We shall prove \eqref{argmax} by induction.  First note that 
\begin{align}
\nonumber
g_1 &= 0,\\
\label{g and e}
g_k &= \alpha e_{k-1},\qquad k=2,3,\ldots,
\intertext{and}
\label{e and s}
e_k s_k^T &= 1,\qquad\quad k = 1,2,\ldots.
\end{align}
Also recall that
$s_k = f_k + g_k + e_k$
and 
\begin{gather}
\label{e and pi}
e_k\,\Pi_{k-1} = e_k.
\end{gather}
For $k=2,$ we trivially have
\[
\|g_2 \,\Pi_1\|^2 = \alpha^2 e_1\left(I - \frac{s_1^Ts_1}{s_1 s_1^T} \right)e_1^T
= \alpha^2 \left(1 - \frac{1}{s_1 s_1^T} \right)
= \alpha^2 \left(1 - \frac{1}{\det(S_1 S_1^T)} \right),
\]
which establishes \eqref{argmax}.
Further, from \eqref{j1} and \eqref{two max}, we can check that 
\begin{align*}
J_2 &= \max_{f_1} \left[ \log (s_{1} s_{1}^T) + 
\log \left(1 + \left(\sqrt{P_2} +|\alpha| \sqrt{1 - \frac{1}{s_1 s_1^T}}\right)^2
\right) \right]\\
&= J_1 + 
\log \left(1 + \left(\sqrt{P_2} + |\alpha| \sqrt{1 - \frac{ 1}{e^{J_1}}} \right)^2\right).
\end{align*}

Now suppose \eqref{argmax} holds for $k = 2, \ldots, m-1$.  For $k \ge
3,$ we observe that
\begin{align*}
\Pi_{k-1} &=
I - S_{k-1}^T (S_{k-1} S_{k-1}^T)^{-1} S_{k-1}\\
&=
I - \left[
\begin{array}{c}
S_{k-2}\\[3pt]
s_{k-1}
\end{array}
\right]^T \left[
\begin{array}{cc}
S_{k-2}S_{k-2}^T & S_{k-2}s_{k-1}^T\\[3pt]
s_{k-1}S_{k-2}^T & s_{k-1}s_{k-1}^T
\end{array}
\right]^{-1} 
\left[
\begin{array}{c}
S_{k-2}\\[3pt]
s_{k-1}
\end{array}
\right]\\
&= I - S_{k-2}^T \left(S_{k-2} S_{k-2}^T\right)^{-1} S_{k-2} - 
\Pi_{k-2}\,{s}_{k-1}^T \left({s}_{k-1}\,\Pi_{k-2}\,{s}_{k-1}^T\right)^{-1}\!{s}_{k-1}\,\Pi_{k-2}\\
&= \Pi_{k-2} (I - \Pi_{k-2}\,{s}_{k-1}^T 
\left({s}_{k-1}\,\Pi_{k-2}\,{s}_{k-1}^T\right)^{-1}\!{s}_{k-1}\,\Pi_{k-2}) \Pi_{k-2}.
\end{align*}
Now from \eqref{g and e}, (\ref{e and s}), and (\ref{e and pi}), we
have
\begin{align}
\|g_k\,\Pi_{k-1}\|^2 &=
g_k\,\Pi_{k-1} g_k^T\notag\\
&= g_k \Pi_{k-2} (I - \Pi_{k-2}\,{s}_{k-1}^T 
\left({s}_{k-1}\,\Pi_{k-2}\,{s}_{k-1}^T\right)^{-1}\!{s}_{k-1}\,\Pi_{k-2})
\Pi_{k-2} g_k^T\notag\\
&= \alpha^2 e_{k-1} \left(I - \Pi_{k-2}\,{s}_{k-1}^T 
\left({s}_{k-1}\,\Pi_{k-2}\,{s}_{k-1}^T\right)^{-1}\!{s}_{k-1}\,\Pi_{k-2}\right)
e_{k-1}^T\notag\\
&= \alpha^2 \left (1 - \frac{1}{ s_{k-1}\,\Pi_{k-2}\,s_{k-1}^T} \right)\notag\\
&= \alpha^2 \left (1 - \frac{1}{ 1 + (f_{k-1}+g_{k-1})\,\Pi_{k-2}\,(f_{k-1} + g_{k-1})^T}
\right).
\label{g pi}
\end{align}
It follows from \eqref{eq orth} -- (\ref{maximizer})
and (\ref{g pi}) that, for fixed $(f_1, \ldots, f_{m-2})$, both
$\det(S_{m-1}S_{m-1}^T)$ and $\|g_m\,\Pi_{m-1}\|$ have the same
maximizer
\begin{align*}
f_{m-1}^* = \sqrt{P_{m-1}} \frac{g_{m-1}\, \Pi_{m-2}}{\|g_{m-1} \,\Pi_{m-2} \|}.
\end{align*}
Plugging this back to \eqref{g pi}, for fixed $(f_1, \ldots, f_{m-2})$,
we have
\[
\max_{f_{m-1}} \|g_m\,\Pi_{m-1}\|^2 = \alpha^2 \left(1 - \frac{1}{1 + 
(\sqrt{P_{m-1}} + \|g_{m-1} \,\Pi_{m-2} \|)^2}\right)
\]
while
\[
\max_{f_{m-1}} \det(S_{m-1} S_{m-1}^T) = 
\det (S_{m-2} S_{m-2}^T) \cdot                                 
\left(1 + \left(\sqrt{P_{m-1}} + \| g_{m-1} \,\Pi_{m-2} \|\right)^2\right).
\]
But from the induction hypothesis, $\det (S_{m-2}S_{m-2}^T)$ and
$\|g_{m-1} \,\Pi_{m-2} \|$ have the same maximizer
$(f_1^*,\ldots,f_{m-2}^*)$.  Thus $\det (S_{m-1}S_{m-1}^T)$ and
$\|g_{m} \,\Pi_{m-1} \|$ have the same maximizer
$(f_1^*,\ldots,f_{m-1}^*)$.  Therefore, we have established
\eqref{argmax} for $k = m$ and hence for all $k = 2,3, \ldots.$ From
\eqref{two max} and \eqref{argmax}, we easily get the desired
recursion formula as
\begin{align*}
J_k &= J_{k-1} + 
\log \left(1 + \left(\sqrt{P_{k}} + 
|\alpha| \sqrt{1 - \frac{ 1}{e^{J_{k-1} - J_{k-2}}}} \right)^2\right),
\qquad k = 2, 3,\ldots.
\end{align*}

\vskip 1em
{\it Step 3. Optimal power allocation over time.}

In the previous step, we solved the optimization problem (\ref{opt 4})
under a fixed power allocation $(P_1,\ldots, P_n)$.  Thanks to the
special structure of the MA(1) noise process, this brute force
optimization was tractable via backward dynamic programming.  Here we
optimize the power allocation $(P_1,\ldots,P_n)$ under the constraint
$\sum_{i=1}^n P_i \le nP$,

As we saw earlier, when $\alpha = 0$, we can use the concavity of 
the logarithm to show that,
for all $n$, 
\begin{align*}
C_{n,\text{FB}} = \frac{1}{2n} J_n(P_1,\ldots,P_n) = \max_{P_i:
\sum_i P_i \le nP} \frac{1}{2n} \sum_{i=1}^n \log (1 + P_i) =
\frac{1}{2} \log (1 + P),
\end{align*}
with $P_1^* = \cdots = P_n^* = P.$ When $\alpha \ne 0$, it is not
tractable to optimize $(P_1,\ldots,P_n)$ for $J_n$ in
(\ref{recursion1}) -- (\ref{recursion3})
to get a closed-form solution of $C_{n,\text{FB}}$ for finite $n$.
The following lemma, however, enables us to figure out the asymptotically
optimal power allocation and to obtain a closed-form
solution for $C_{\text{FB}} = \lim_n C_{n,\text{FB}}$.

\begin{lemma}
\label{lemma asymp}
Let $\psi: [0,\infty) \times [0,\infty) \to [0,\infty)$ such that
the following conditions hold:
\setenumerate{topsep=0em, partopsep=0em}
\begin{enumerate}[label=(\roman*), leftmargin=*]
\item $\psi(\xi,\zeta)$ is continuous, concave in $(\xi,\zeta)$,
and strictly concave in $\xi$ for all $\zeta > 0$;
\item $\psi(\xi,\zeta)$ is increasing in $\xi$ and $\zeta$,
respectively; and
\item for each $\zeta > 0$, there is a unique solution
$\xi^*(\zeta) > 0$ to the equation $\xi = \psi(\xi,\zeta).$
\end{enumerate}
For some fixed $P > 0$, let $\{P_i\}_{i=1}^\infty$ be any infinite
sequence of nonnegative numbers satisfying
\begin{align*}
\limsup_{n\to\infty} \frac{1}{n}\sum_{i=1}^n P_i \le P.
\end{align*}
Let $\{\xi_i\}_{i=0}^\infty$ be defined recursively as
\begin{align*}
\xi_0 &= 0,\\
\xi_i &= \psi(\xi_{i-1}, P_i),\qquad i = 1,2,\ldots.
\end{align*}
Then
\begin{align*}
\limsup_{n\to\infty} \frac{1}{n} \sum_{i=1}^n \xi_i \le \xi^*,
\end{align*}
where $\xi^* = \xi^*(P)$ is the unique solution to $\xi = \psi(\xi,
P)$.  Furthermore, if $P_i \equiv P,$ $i = 1,2,\ldots,$ then the
corresponding $\xi_i$ converges to $\xi^*$.
\end{lemma}

\begin{figure}
\begin{center}
\input{conv.pstex_t}
\caption{Convergence to the unique point $\xi^*$.}
\end{center}
\end{figure}

\begin{proof}
Fix $\epsilon > 0$. From the concavity and monotonicity of 
$\psi$, for $n$ sufficiently large,
\begin{align*}
\frac{1}{n} \sum_{i=1}^n \xi_i &= \frac{1}{n} \sum_{i=1}^n
\psi(\xi_{i-1}, P_i)\\
&\le \psi\left(\frac{1}{n} \sum_{i=1}^n
\xi_{i-1}, \frac{1}{n} \sum_{i=1}^n P_i\right)\\
&\le
\psi\left(\frac{1}{n} \sum_{i=1}^n \xi_{i-1}, P + \epsilon\right).
\end{align*}
Taking $\limsup$ on both sides and using the continuity of $\psi$, we have
\begin{align*}
\xi^{**} := \limsup_{n\to\infty} \frac{1}{n} \sum_{i=1}^n \xi_i 
\le \limsup_{n\to\infty} \psi\left(\frac{1}{n} \sum_{i=1}^n \xi_{i-1}, P + \epsilon \right) =
\psi(\xi^{**}, P + \epsilon).
\end{align*}
Since $\epsilon$ is arbitrary and $\psi$ is continuous, we have $
\xi^{**} \le \psi(\xi^{**}, P).  $ But from uniqueness of $\xi^*$ and
strict concavity of $\psi$ in $\xi$, we have
\begin{align}
\label{psi}
\xi \le \xi^* \text{ if and only if } \xi \le \psi(\xi, P).
\end{align}
Thus $\xi^{**} \le \xi^*.$

It remains to show that we can actually attain $\xi^*$ by choosing
$P_i \equiv P$, $i = 1,2,\ldots.$ Let $\xi_i = \psi(\xi_{i-1}, P), i =
1,2,\ldots.$ From the monotonicity of $\psi(\cdot, P)$ and
(\ref{psi}), we have
\begin{align*}
\xi_{i-1} ~\le~ \xi_{i} = \psi(\xi_{i-1},P) ~\le~ \xi^* = \psi(\xi^*, P),\qquad i = 1,2,\ldots.
\end{align*}
Thus the sequence $\{\xi_i\}$ has a limit, which we denote as
$\xi^{**}$.  But from the continuity of $\psi(\cdot, P)$,
 we must have
\begin{align*}
\xi^{**} = \lim_{n\to\infty} \xi_n = \lim_{n\to\infty} \psi(\xi_n ,P)
= \psi\left(\,\lim_{n\to\infty} \xi_n, P\right) = \psi(\xi^{**}, P).
\end{align*}
Thus $\xi^{**} = \xi^{*}$.
\end{proof}

We continue our main discussion.  Define
\begin{align*}
\psi(\xi,\zeta) &:= 
\frac{1}{2} \log \left(1 + \left(\sqrt{\zeta} + 
|\alpha| \sqrt{1 - \frac{ 1}{e^{2\xi}}} \right)^2\right).
\end{align*}
The conditions (i) -- (iii) of Lemma 2 can be easily checked.  For
concavity, we rely on the simple composition rule for concave
functions~\cite[Section 3.2.4]{Boyd 2004} without messy calculus.  Let
$\psi_1(\xi) = \frac{1}{2} \log (1 + \xi)$, $\psi_2(\xi,\zeta) =
(\sqrt{\xi} + \sqrt{\zeta})^2,$ and $\psi_3(\xi) = |\alpha|^2 (1 -
\exp(-2\xi)).$ Then $\psi(\xi,\zeta) = \psi_1(\psi_2(\psi_3(\xi),
\zeta)).$ Now that $\psi_1$ is strictly concave and strictly
increasing, $\psi_2$ is concave (strictly concave in $\xi$ alone for
each $\zeta > 0$) and elementwise strictly increasing, and $\psi_3$ is
strictly concave, we can conclude that $\psi$ is concave in
$(\xi,\zeta)$ and strictly concave in $\xi$ for all $\zeta > 0$.
Since for any $\zeta > 0$, $\psi(0,\zeta) > 0$ and $\psi(\xi,\zeta)
\to c(\zeta) < \infty$ as $\xi$ tends to infinity, the uniqueness of
the root of $\xi = \psi(\xi,\zeta)$ is trivial from the continuity of
$\psi$.

For an arbitrary infinite sequence $\{P_i\}_{i=1}^\infty$ satisfying
\begin{align}
\limsup_{n\to\infty} \frac{1}{n} \sum_{i=1}^n P_i \le n P,
\label{asymp power}
\end{align}
we define
\begin{align*}
\xi_0 &= 0,\\
\xi_i &= \psi(\xi_{i-1}, P_i),\qquad i = 1,2,\ldots.
\end{align*}
Note that 
\begin{align*}
\xi_1 &= \frac{1}{2} J_1(P_1),\\ \xi_i &= \frac{1}{2}
\left(J_i(P_1,\ldots,P_i) - J_{i-1}(P_1,\ldots,P_{i-1})\right),\qquad
i = 2,3,\ldots.
\end{align*}
Now from Lemma~2, we have 
\begin{align*}
\limsup_{n\to\infty} \frac{1}{2n} J_n(P_1,\ldots,P_n) 
= \limsup_{n\to\infty} \frac{1}{n} \sum_{i=1}^n \xi_i \le \xi^*,
\end{align*}
where $\xi^*$ is the unique solution to
\begin{align*}
\xi = \psi(\xi, P) = 
\frac{1}{2} \log \left(1 + \left(\sqrt{P} + 
|\alpha| \sqrt{1 - \frac{ 1}{e^{2\xi}}} \right)^2\right).
\end{align*}
Since our choice of $\{P_i\}$ is arbitrary, we conclude that
\begin{align*}
\sup\,\limsup_{n\to\infty} \frac{1}{2n} J_n(P_1,\ldots,P_n) = 
\lim_{n\to\infty} \frac{1}{2n} J_n(P,\ldots,P) = 
\xi^*,
\end{align*}
where the supremum (in fact, maximum) is over all infinite sequences
$\{P_i\}$ satisfying the asymptotic average power
constraint~(\ref{asymp power}).

Finally, we prove that $C_\text{FB} = \xi^*$.  More specifically, we
will show that
\begin{align}
C_\text{FB} &= \lim_{n\to\infty} C_{n,\text{FB}}\notag\\
&= \lim_{n\to\infty} \max_{P_i: \sum_i P_i \le nP}
\frac{1}{2n} J_n(P_1,\ldots, P_n)\label{lim sup}\\
&= \sup_{\{P_i\}_{i=1}^\infty}\,\limsup_{n\to\infty}
\frac{1}{2n}J_n(P_1,\ldots,P_n)
\label{sup lim}\\
&= \xi^*.\notag
\end{align}
The only subtlety here is how to justify the interchange of the order of
limit and supremum in (\ref{lim sup}) and (\ref{sup lim}).
It is easy to verify that 
\begin{align*}
\lim_{n\to\infty} \max_{P_i: \sum_i P_i \le nP}
\frac{1}{2n} J_n(P_1,\ldots, P_n) 
&\ge \sup_{\{P_i\}_{i=1}^\infty}\,\limsup_{n\to\infty}
\frac{1}{2n}J_n(P_1,\ldots,P_n),
\end{align*}
for it is always advantageous to choose for each $n$ a finite sequence
$(P_1,\ldots,P_n)$ with $\sum_{i=1}^n P_i \le nP$ for each $n$ rather
than fixing a single infinite sequence $\{P_i\}$ with $P_i = P$ for
all $i$.  (Recall that the supremum on the right side is achieved by
the uniform power allocation.)

To prove the other direction of inequality, we fix $\epsilon>0$ and
choose $n$ and $(P_1^*,\ldots,P_n^*)$ such that
\begin{gather}
\sum_{i=1}^n P_i^* \le nP\notag\\
\intertext{and}
\frac{1}{2n}
J_n(P_1^*,\ldots, P_n^*) \ge C_{\text{FB}} - \epsilon.
\label{j_n q}
\end{gather}
Now we construct an infinite sequence $\{P_i\}_{i=1}^\infty$ by
concatenating $(P_1^*,\ldots,P_n^*)$ repeatedly,
that is, $P_{kn + i} = P_i^*$ for all $i = 1,\ldots, n,$ and $k = 0,
1, \ldots.$ Obviously, this choice of $\{P_i\}$ satisfies the power
constraint (\ref{asymp power}).  As before, let $\xi_{i} =
\psi(\xi_{i-1}, P_i),$ $i = 1,2,\ldots.$ By induction, it is easy to
see that
\begin{align}
\label{xi ineq}
\xi_i \le \xi_{kn+i},\qquad i = 1,2,\ldots, n
\end{align}
for all $k = 0,1,\ldots.$ For $k=0$, \eqref{xi ineq} holds trivially.
Suppose \eqref{xi ineq} holds for $k = 0,\ldots, m-1.$ Then from the
monotonicity of $\psi(\xi,\zeta)$ in $\xi$, 
we have
\begin{align*}
\xi_{1} = \psi(\xi_{0}, P_1) = \psi(\xi_0, P_1^*) &
\le \psi(\xi_{mn}, P_1^*) =
\psi(\xi_{mn}, P_{mn+1}) = \xi_{mn+1},\\
\xi_{2} = \psi(\xi_{1}, P_2) = \psi(\xi_1, P_2^*) 
&\le \psi(\xi_{mn+1}, P_2^*) =
\psi(\xi_{mn+1}, P_{mn+2}) = \xi_{mn+2},
\intertext{and in general}
\xi_{i} = \psi(\xi_{i-1}, P_i) = \psi(\xi_{i-1}, P_i^*) 
&\le \psi(\xi_{mn+i-1}, P_i^*) =
\psi(\xi_{mn+i-1}, P_{mn+i}) = \xi_{mn+i}
\end{align*}
for all $i = 1,\ldots, n$.  Thus, \eqref{xi ineq} holds for all $k$.
Therefore
\begin{align*}
\frac{1}{2kn} J_{kn}(P_1,\ldots,P_{kn}) 
= \frac{1}{kn} \sum_{i=1}^{kn} \xi_i
\ge \frac{1}{kn} \left(k\cdot\sum_{i=1}^{n} \xi_i \right)
= \frac{1}{2n} J_{n}(P_1,\ldots,P_n).
\end{align*}
which, combined with (\ref{j_n q}), implies that
\begin{gather*}
\limsup_{n\to\infty} \frac{1}{2n} J_n(P_1,\ldots,P_n) \ge
C_{\text{FB}} - \epsilon,\quad k = 1,2,\ldots,
\intertext{which, in turn, implies that}
\sup_{\{P_i\}_{i=1}^\infty} \limsup_{n\to\infty} \frac{1}{2n}
J_n(P_1,\ldots,P_n) \ge C_{\text{FB}} - \epsilon.
\end{gather*}
Since $\epsilon$ is arbitrary, we have the desired inequality. Thus
$C_\text{FB} = \xi^*.$

We conclude this section by characterizing the capacity $C_\text{FB} =
\xi^*$ in an alternative form.  Recall that $\xi^*$ is the unique
solution to
\begin{align*}
\xi = 
\frac{1}{2} \log \left(1 + \left(\sqrt{P} + 
|\alpha| \sqrt{1 - \frac{ 1}{e^{2\xi}}} \right)^2\right).
\end{align*}
Let $x_0 = \exp(-\xi^*),$ or equivalently, $\xi^* = - \log x_0$.
It is easy to verify that $0 < x_0 \le 1$ is the unique positive
solution to
\begin{align*}
\frac{1}{x^2} = 1 + \left(\sqrt{P} + |\alpha| \sqrt{1-x^2}\right)^2,\notag
\intertext{or equivalently,}
P\,x^2 = (1-x^2)(1-|\alpha|x)^2.
\end{align*}
This establishes the feedback capacity $C_\text{FB}$ of the additive
Gaussian noise channel with the noise covariance $K'_Z$, which is, in
turn, the feedback capacity of the first-order moving average additive
Gaussian noise channel with parameter $\alpha$, as is argued at the
end of Step 1 and proved in the Appendix.  This completes the proof of
Theorem 1.

\section{Discussion}
\label{sec-disc}
The derived asymptotically optimal feedback input signal sequence, or
equivalently, the (sequence of) matrices $(K_V^*{(n)}, B^*{(n)})$ has
two prominent properties.  First, the optimal $(K_V^*(n),B^*(n))$ for
the $n$-block can be found sequentially, built on the optimal
$(K_V^*(n-1), B^*(n-1))$ for the $(n-1)$-block.  Although this
property may sound quite natural, it is not true in general for other
channel models.  Later in this section, we will see an MA(2) channel
counterexample.  As a corollary to this sequentiality property, the
optimal $K_V$ has rank one, which agrees with the previous result by
Ordentlich~\cite{Ordentlich 1995a}.  Secondly, the current input
signal $X_k$ is orthogonal to the past output signals $(Y_1,\ldots,
Y_{k-1})$.  In the notation of Section 3, we have $f_k S_{k-1}^T = 0.$
This orthogonality property is indeed a necessary condition for the
optimal $(K_V^*, B^*)$ for any (possibly nonstationary nonergodic)
noise covariance matrix $K_Z$~\cite{Ihara 1979, Ordentlich 1995a}.  It
should be pointed out that the recursion formula \eqref{recursion1} --
\eqref{recursion3} can be also derived from the orthogonality property
and the optimality of rank-one $K_V$.

We explore the possibility of extending the current proof technique to
a more general class of noise processes.  The immediate answer is
negative.  We comment on two simple cases: MA(2) and AR(1).  Consider
the following MA(2) noise process which is essentially two interleaved
MA(1) processes:
\begin{align*}
Z_i = U_i + \alpha U_{i-2}, \qquad i = 1,2,\ldots.
\end{align*}
It is easy to see that this channel has the same capacity as the MA(1)
channel with parameter $\alpha$, which can be attained by signalling
separately for each interleaved MA(1) channel.  This suggests that the
sequentiality property does not hold for this example.  Indeed, if we
sequentially optimize the $n$-block capacity, we achieve the rate $-
\log x_0$, where $x_0$ is the unique positive root of the sixth order
polynomial
\begin{align*}
P\,x^2 = (1-x^2) (1 - |\alpha| x^2)^2.
\end{align*}
It is not difficult to see that this rate is strictly less than the
feedback capacity of the interleaved MA(1) channel unless $\alpha =
0$.  A similar argument can prove that Butman's conjecture on the
AR($k$) capacity~\cite[Abstract]{Butman 1976} is not true in general
for $k > 1$.

In contrast to MA(1) channels, we are missing two basic ingredients
for AR(1) channels --- the optimality of rank-one $K_V$ and the
asymptotic optimality of the uniform power allocation.  Under these
two conditions, both of which are yet to be justified, it is
known~\cite{Wolfowitz 1975, Tiernan 1976} that the optimal achievable
rate is given by $ - \log x_0,$ where $x_0$ is the unique positive
root of the fourth order polynomial
\begin{align*}
P\,x^2 = \frac{1-x^2}{(1 + |\alpha| x^2)^2}.
\end{align*}
There is, however, a major difficulty in establishing the above two
conditions by the two-stage optimization strategy we used in the
previous section, namely, first maximizing $(f_1,\ldots,f_n)$ and then
$(P_1,\ldots,P_n)$.  For certain values of individual signal power
constraints $(P_1,\ldots,P_n)$, the optimal $(f_1,\ldots,f_n)$ does
not satisfy the sequentiality, resulting in $K_V$ with rank higher
than one.  Hence, a greedy maximization of $\log\det(S_kS_k^T)$ does
not establish the recursion formula for the AR(1) $n$-block capacity
that corresponds to our (\ref{recursion1}) -- (\ref{recursion3}):
\begin{align*}
J_0 &:= 0\\
J_1 &= \log (1+P_1)\\
J_{k+1} &= J_k + 
\log \left(1 + \left(\sqrt{P_{k+1}}+
|\alpha|\sqrt{P_k}e^{-\frac{J_k - J_{k-1}}{2}}\right)^2 \right),
\quad k=1, 2,\ldots.
\end{align*}
(See \cite{Wolfowitz 1975, Tiernan 1976, Butman 1976} for the
derivation of the above recursion formula.)  Even under the assumption
that the optimal $K_V$ for the AR(1) channel has rank one, it has been
unclear whether the uniform power allocation over time is
asymptotically optimal.  

Nonetheless, using a technique similar to the one deployed in Lemma~2,
we can prove the optimality of the uniform power allocation, resolving
a question raised by Butman~\cite{Butman 1969, Butman 1976} and
Tiernan~\cite{Tiernan 1976} among others.  Since the proof is a little
technical in nature, we defer it to the Appendix.

Finally we show that the feedback capacity of the MA(1) channel can be
achieved by using a simple stationary filter of the noise innovation
process.  Before we proceed, we point out that the optimal input
process $\{X_i\}$ we obtained in the previous section is
asymptotically stationary.  This observation is not hard to prove
through the well-developed theory on the asymptotic behavior of
recursive estimators~\cite[Chapter 14]{Kailath 2000}.

At the beginning, we send\footnote{ Technically, we generate $2^{nR}\;
X_1(W)$ code functions i.i.d.\ according to $N(0,P)$ for some $R <
C_{\text{FB}}$, and transmit one of them.}
\begin{align*}
X_1 \sim N(0,P).
\end{align*}
For subsequent transmissions, we
transmit the filtered version of the noise innovation process up to
the time $k-1$:
\begin{align}
X_k &= \beta\, X_{k-1} + \sigma U_{k-1},\qquad k = 2,3,\ldots.
\label{stat trans}
\end{align}
In other words, we use a first-order
regressive filter with transfer function given by
\begin{align}
\label{opt filter}
\frac{\sigma z^{-1}}{1 - \beta z^{-1}}.
\end{align}
Here $\beta = -\sgn(\alpha)\,x_0$ with $x_0$ being the same unique
positive root of the fourth-order polynomial (\ref{ma1 cap}) in Theorem 1.
The scaling factor $\sigma$ is chosen to satisfy the power constraint
as
\begin{align*}
\sigma =  \sgn(\alpha) \sqrt{P\,(1-\beta^2)},
\end{align*}
where
\begin{align*}
\sgn(\zeta) = \left\{
\begin{array}{ll}
1,&\quad \zeta \ge 0,\\
-1,&\quad \zeta < 0.
\end{array}\right.
\end{align*}
This input process and the MA(1) noise process
\begin{align*}
Z_k &= \alpha U_{k-1} + U_k,\qquad k = 1,2,\ldots,
\end{align*}
yield the output process given by
\begin{align*}
Y_1 &= X_1 + \alpha U_0 + U_1,\\[3pt]
Y_k  &= \beta\, X_{k-1} + (\alpha + \sigma)
U_{k-1} + U_k,\\
&= \beta\; Y_{k-1} -\alpha\beta\,U_{k-2} + (\alpha -
\beta + \sigma)U_{k-1},\qquad k=2,3,\ldots,
\end{align*}
which is asymptotically stationary
with power spectral density
\begin{align}
\nonumber
S_Y(\omega) &=
\left\arrowvert 1 + \alpha e^{-j\omega} + 
\frac{\sigma e^{-j\omega}}{1 - \beta e^{-j\omega}}
\right\arrowvert^2 \\
\nonumber
&= 
\left\arrowvert \frac{1 + (\alpha - \beta + \sigma) e^{-j\omega}
 - \alpha\beta e^{-j2\omega}}
{(1 - \beta e^{-j\omega})} \right\arrowvert^2\\
\label{spec zero}
&= 
\left\arrowvert
\frac{(1 + \alpha\beta^2 e^{-j\omega})
(1 - \beta^{-1}e^{-j\omega})}
{(1 - \beta e^{-j\omega})} \right\arrowvert^2\\
\nonumber
&= \beta^{-2} | 1 + \alpha\beta^2 e^{-j\omega} |^2.
\end{align}
The ``asymptotic stationarity'' here should not bother us since
$\{Y_k\}$ is stationary for $k\ge2$ and $h(Y_1 | Y_2,\ldots, Y_n)$ is
uniformly bounded in $n$; hence the entropy rate of the process
$\{Y_k\}_{k=1}^\infty$ is determined by $(Y_2, Y_3,\ldots)$.  Thus
from (\ref{ent rate}) in Section~\ref{sec-noise}, the entropy rate of
the output process $\{Y_k\}$ is given by
\begin{align*}
\frac{1}{4\pi} \int_{-\pi}^{\pi} \log \left(2\pi e S_Y(\omega)\right) d\omega
= \frac{1}{2} \log (2\pi e \beta^{-2}) = \frac{1}{2} \log (2\pi e x_0^{-2}). 
\end{align*}
Hence we attain the feedback capacity $C_\text{FB}$.
Furthermore, it can be shown that the mean-square error of $X_1$ given
the observations $Y_1,\ldots,Y_n$ decays 
exponentially with rate $\beta^{-2} = 2^{2C_\text{FB}}$.
In other words,
\begin{align}
\var(X_1|Y_1,\ldots,Y_n)
= E (X_1 - E(X_1|Y_1,\ldots,Y_n))^2 \doteq P\;2^{-2nC_\text{FB}}.
\label{error decay}
\end{align}

Note that the optimal filter \eqref{opt filter} has an interesting
feature.  In the light of \eqref{spec zero}, we can think of the
output process $\{Y_k\}$ as the filtered version of the noise
innovation process $\{U_k\}$ through the monic filter
\[
1-\alpha z^{-1} + \frac{\sigma z^{-1}}{1-\beta z^{-1}} = 
\frac{(1+\alpha\beta^2 z^{-1})(1-\beta^{-1}z^{-1})}
{1- \beta z^{-1}}.
\]
As the entropy rate formula~\eqref{ent rate}, or more fundamentally,
Jensen's formula~\eqref{jensen} shows, the entropy rate of $\{Y_k\}$
is totally determined by all zeros of the filter outside the unit
circle, which, for our case, is $\beta^{-1}$.  Hence, we can interpret
the feedback capacity problem as the problem of relocating the zero of
the original noise filter $1 + \alpha z^{-1}$ to the outside of the
unit circle and making the modulus of that zero as large as possible
by adding a causal filter $H(z)$ using the power $(2\pi)^{-1} \int
|H(e^{-j\omega})|^2 d\omega = P$.  Here we have shown that the optimal
filter is given by \eqref{opt filter}.  Under this interpretation, the
initial input $X_1$ is merely a perturbation which guarantees that the
output process is not causally invertible from the innovation process
and hence that the entropy rate is fully determined by the spectral
density of the stationary part.  (Without $X_1$, the entropy rate of
$\{Y_k\}$ is exactly same as the entropy rate of $\{Z_k\}.$)

From a classical viewpoint, we can interpret the signal $X_k$ as the
adjustment of the receiver's estimate of the message-bearing signal
$X_1$ after observing $(Y_1, \ldots, Y_{k-1})$.  We can further check
that following signalling schemes are equivalent (and thus optimal) up
to scaling:
\begin{align*}
X_k \quad&\propto\quad X_1 - \hat{X}_1(Y^{k-1})\\[.25em]
&\propto\quad X_j - \hat{X}_j(Y^{k-1})\qquad\qquad (j < k)\\[.25em]
&\propto\quad U_{k-1} - \hat{U}_{k-1}(Y^{k-1})\\[.25em]
&\propto\quad \hat{Z}_k(Y^{k-1},X^{k-1}) - \hat{Z}_k(Y^{k-1}).
\end{align*}
The connection to the
Schalkwijk-Kailath coding scheme is now apparent.  Recall that there
is a simple linear relationship~\cite[Section 3.4]{Kailath
2000}~\cite[Section 4.5]{Luenberger 1969} between the minimum mean
square error estimate (in other words, the minimum variance biased
estimate) for the Gaussian input $X_1$ and the maximum likelihood
estimate (or equivalently, the minimum variance unbiased estimate) for
an arbitrary real input $\theta$.  Thus we can easily transform the
above coding scheme based on the asymptotic equipartition
property~\cite{CP 1989} to a variant of the Schalkwijk-Kailath linear
coding scheme based on the maximum likelihood nearest neighborhood
decoding of uniformly spaced $2^{nR}$ points.  More specifically, we
send as $X_1$ one of $2^{nR}$ possible signals, say, $\theta \in
\Theta := \{-\sqrt{P}, -\sqrt{P} + \Delta, -\sqrt{P} + 2\Delta,
\ldots, \sqrt{P} - 2\Delta, \sqrt{P} - \Delta, \sqrt{P}\}$, where
$\Delta = \frac{2\sqrt{P}}{2^{nR} - 1}$.  Subsequent transmissions follow
(\ref{stat trans}).  The receiver forms the maximum likelihood
estimate $\hat{\theta}_n(Y_1, \ldots,Y_n)$ and finds the nearest
signal point to $\hat\theta_n$ in $\Theta$.

The analysis of the error for this coding scheme follows
Schalkwijk~\cite{Schalkwijk 1966} and Butman~\cite{Butman 1969}.
From~(\ref{error decay}) and the standard result on the relationship
between the minimum variance unbiased and biased estimation errors,
the maximum likelihood estimation error $\hat{\theta}_n - \theta$ is,
conditioned on $\theta$, Gaussian with mean $\theta$ and variance
exponentially decaying with rate $\beta^{-2} = 2^{2nC_\text{FB}}$.
Thus, the nearest neighbor decoding error, ignoring lower order terms,
is given by
\begin{gather*}
P_e^{(n)} = E_\theta \bigg[\Pr\Big( |\hat\theta_n - \theta| \ge 
\frac{\Delta}{2}
\:\Big|\: \theta\Big)\bigg]
\doteq \erfc\Big(\sqrt{\frac{3}{2\sigma_\theta^2}}
2^{n(C_\text{FB} - R)}\Big),
\intertext{where}
\erfc(x) = \frac{2}{\sqrt{\pi}} \int_{x}^\infty \exp(-t^2) dt,
\end{gather*}
and $\sigma_\theta^2$ is the variance of input signal $\theta$ chosen
uniformly over $\Theta$.  As far as $R < C_\text{FB}$, the decoding
error decays doubly exponentially in $n$.  Note that this coding
scheme uses only the second moments of the noise process.  This implies that
the rate $C_\text{FB}$ is achievable for the additive noise channel
with any non-Gaussian noise process with the same covariance matrix.

\section*{Appendix}

{\bf Asymptotic equivalence of $K_Z$ and $K'_Z$ for feedback
capacity}

Recall that $Z^n \sim N_n(0, K_Z)$ and $\tilde{Z}^n \sim N_n(0,K'_Z)$.
To stress the dependence of the capacity on the power constraint and
the noise covariance, we use the notation $C_{n,\text{FB}}(K, P)$ for
$n$-block feedback capacity of the channel with $n$-block noise
covariance matrix $K$ under the power constraint $E \sum_{i=1}^n X_i^2
\le nP.$ With a little abuse of notation, we similarly use
$C_\text{FB}(K, P)$ for feedback capacity of the channel with infinite
noise covariance matrix naturally extended from $K$ under the power
constraint $P$.

Suppose $(B^*, K_V^*)$ maximizes
\begin{align*}
C_{n,\text{FB}}(K_Z,P) = \max \frac{1}{2n} \log \frac{
\det((B+I)K_Z(B+I)^T + K_V)}
{\det(K_Z)}
\end{align*}
and $(B^{**}, K_V^{**})$ maximizes $C_{n,\text{FB}}(K'_Z)$.
Since ${K}'_Z \preceq K_Z$, we have
\begin{align*}
\tr \left( B^{*} K'_Z (B^{*})^T + K_V^{*} \right)
\le \tr \left( B^{*} K_Z (B^{*})^T + K_V^{*} \right) \le nP,
\end{align*}
which shows that $(B^{*}, K_V^{*})$ is a feasible (not necessarily
optimal) solution to $C_{n,\text{FB}}(K_Z',P)$.  On the other hand, we
have
\begin{align}
(B^*+I){K}'_Z(B^*+I)^T \preceq (B^*+I){K}_Z(B^*+I)^T,
\label{noise covs}
\end{align}
{so that}
\begin{align}
C_{n,\text{FB}}(K_Z,P) &= I(V^n; V^n + (B^* + I) Z^n)|_{V^n \sim
N(0,K_V^*)}\notag\\ &\le I(V^n; V^n + (B^* + I) \tilde{Z}^n)|_{V^n
\sim N(0,K_V^*)}\label{data proc}\\ &\le I(V^n; V^n + (B^{**} + I)
\tilde{Z}^n)|_{V^n \sim N(0,K_V^{**})}\label{optimality}\\ &=
C_{n,\text{FB}}({K}'_Z,P),\notag
\end{align}
where (\ref{data proc}) follows from (\ref{noise covs}), divisibility
of the Gaussian distribution, and the data processing
inequality~\cite[Section 2.8]{CT 1991}; and \eqref{optimality} follows
from the optimality of $(B^{**}, K_V^{**})$ for
$C_{n,\text{FB}}(K_Z',P)$ and the feasibility of $(B^{*}, K_V^{*})$
for $C_{n,\text{FB}}(K_Z',P)$.  By letting $n$ tend to infinity, we
obtain
\begin{equation}
\lim_{n\to\infty} C_{n,\text{FB}}(K_Z,P) \le
\liminf_{n\to\infty} C_{n,\text{FB}}(K'_Z,P).
\label{one direction}
\end{equation}

For the other direction of inequality, we first consider the case
$|\alpha| < 1 $.  Fix $n$ and define the conditional covariance matrix
$K_Z^{(m)}$, $m = 0,1,\ldots,$ of $Z^n$ conditioned on $m$ past values
as
\begin{align*}
K_Z^{(0)} &:= K_Z,\\
K_Z^{(m)} &:= \cov (Z^n | Z_0, \ldots, Z_{-m+1}),\qquad m = 1,2,\ldots.
\end{align*}
It is easy to see that under this notation, the (elementwise) limit of
covariance matrices $K_Z^{(m)}$ exists and
\[
\lim_{m\to\infty} K_Z^{(m)} = K'_Z.
\]
By sending a length-$m$ training sequence over the channel with the
noise covariance matrix $K_Z$, i.e., by transmitting $X_{-m+1} = \cdots =
X_0 = 0$ and then estimating the noise process at the receiver using
$Z_0, \ldots, Z_{-m+1}$, we can achieve the rate
$nC_{n,\text{FB}}(K_Z^{(m)})$ over $n+m$ transmissions.  Hence, we
have
\[
n C_{n,\text{FB}}(K_Z^{(m)},P) \le (n+m)C_{n+m,\text{FB}}(K_Z, P)
\]
for all $P$.  By carefully increasing both $n$ and $m$, we will derive
the desired inequality.

Consider using $(B^{**}, K_V^{**})$, which is optimal for the channel
with noise covariance matrix $K'_Z$, for the channel with noise
covariance $K_Z^{(m)}$.  Since $K_Z^{(m)} \preceq K'_Z$, the resulting
power usage can be greater than $nP$.  However, we have
\begin{align*}
\tr \left(K_V^{**} + B^{**} K_Z^{(m)} (B^{**})^T\right)
&= 
\tr \left(K_V^{**} + B^{**} K'_Z (B^{**})^T 
+ B^{**} (K_Z^{(m)} - K'_Z) (B^{**})^T \right)\\
&\le nP + \tr \left(B^{**} (K_Z^{(m)} - K'_Z) (B^{**})^T\right).
\end{align*}
Now observe that $K_Z^{(m)}$ and $K_Z'$ differ only at the $(1,1)$
entry.  Furthermore, the convergence of $K_Z^{(m)}(1,1) =
\var(Z_1|Z_0, \ldots, Z_{m-1})$ to $K'_Z(1,1) =
\var(Z_1|Z_0,Z_{-1},\ldots)$ is exponentially fast in $m$ (uniformly
in $n$).  Hence, we can bound the amount of additional power usage as
\begin{align*}
\tr \left(B^{**} (K_Z^{(m)} - K'_Z) (B^{**})^T \right) &\le n^2
\max_{1 \le i,j \le n}(B_{ij}^{**})^2 \max_{1 \le i,j \le n}
(K_Z^{(m)} - K'_Z{(m)})\\
&\le c n^3 e^{-m} =: n \epsilon_{n,m},
\end{align*}
where $c$ is a constant independent of $n$ and $m$.  Combining
above observations, we have the following chain of inequalities for
all $n$ and $m$:
\begin{align}
\nonumber
(n+m) C_{n+m,\text{FB}} &(K_Z, P+\epsilon_{n,m}) \\
\nonumber
&\ge n C_{n,\text{FB}} (K_Z^{(m)}, P+\epsilon_{n,m}) \\
\nonumber
&\ge \frac{1}{2} \left 
[\log \det \left(K_V^{**} + (I + B^{**}) K_Z^{(m)} (I + B^{**})^T \right)
-  \log\det K_Z^{(m)} \right]\\
\nonumber
&\ge \frac{1}{2} \left 
[\log \det \left(K_V^{**} + (I + B^{**}) K'_Z (I + B^{**})^T \right)
-  \log\det K_Z^{(m)} \right]\\
\label{final inequality}
&= n C_{n,\text{FB}}(K'_Z, P) +
\frac{1}{2} \left [\log \det K'_Z -  \log\det K_Z^{(m)} \right].
\end{align}
Finally we let $n$ and $m$ grow to infinity such that
\[
\frac{m}{n} \to 0 \quad\text{ and }\quad n^2 e^{-m} \to 0.
\]
The inequality \eqref{final inequality} certainly implies that
\[
\lim_{n\to\infty}C_{n,\text{FB}}(K_Z, P+\epsilon) 
\ge \limsup_{n\to\infty}C_{n,\text{FB}}(K'_Z, P)
\]
for every $\epsilon >0$.  The desired inequality follows from the
continuity of the $C_\text{FB}(K_Z, P)$ in $P$.  

For the case $|\alpha| = 1$, we can perturb the noise process using a
negligible amount of power and proceeds similarly as above.  Indeed,
if we perturb the original covariance matrices $K'_Z$ and $K_Z$ into
the perturbed covariance matrices $K'_Z(\epsilon)$ and $K_Z(\epsilon)$
that correspond to the MA(1) process with parameter
$\alpha(1-\epsilon)$, we have
\begin{align}
\label{perturb1}
\limsup_{n\to\infty} C_{n,\text{FB}}(K'_Z, P)
&\le C_\text{FB}(K'_Z(\epsilon), P + \delta_1(\epsilon))\\
\label{perturb2}
&= C_\text{FB}(K_Z(\epsilon), P + \delta_1(\epsilon))\\
\label{perturb3}
&\le C_\text{FB}\left((1+\delta_2(\epsilon))^{-1}K_Z, 
P+ \delta_3(\epsilon)\right)\\
\nonumber
&= C_\text{FB}\left(K_Z, 
(1+\delta_2(\epsilon))(P+ \delta_3(\epsilon))\right),
\end{align}
where \eqref{perturb1} follows because we can transform the channel
$K'_Z(\epsilon)$ into $K'_Z$ using very small power, \eqref{perturb2}
follows from the result for $|\alpha| < 1$ we obtained above, and
\eqref{perturb3} follows since we can perturb the channel
$(1+\delta_2(\epsilon))^{-1} K_Z$ into $K_Z(\epsilon)$ by adding some
extra white noise.  Since $\delta_k(\epsilon) \to 0$ as $\epsilon \to
0$, $k = 1,2, 3,$ and $C_\text{FB}(K_Z, P)$ is continuous in $P$, we
have 
\[
\limsup_{n\to\infty} C_{n,\text{FB}}(K'_Z, P) \le C_\text{FB}(K_Z, P).
\]
This completes the proof of the asymptotic equivalence of $K_Z$ and
$K'_Z$.  \qed

\vskip 1em
\noindent{\bf Optimality of uniform power allocation for the
Schalkwijk-Kailath-Butman coding scheme for the AR(1) Gaussian
feedback channel.}

Recall that for the AR(1) Gaussian feedback channel, the best
$n$-block achievable rate $R_n$ of the Schalkwijk-Kailath-Butman
coding scheme, or equivalently, the best achievable rate over all
$K_V$ with rank one, is given by
\[
R_n = \max_{P_i:\sum_i P_i \le nP} \frac{1}{2n} J_n(P_1,\ldots,P_n),
\]
where
\begin{align}
\label{ar1 recursion1}
J_0 &:= 0\\
\label{ar1 recursion2}
J_1 &= \log (1+P_1)\\
\label{ar1 recursion3}
J_{k+1} &= J_k + 
\log \left(1 + \left(\sqrt{P_{k+1}}+
|\alpha|\sqrt{P_k}e^{-\frac{J_k - J_{k-1}}{2}}\right)^2 \right),
\quad k=1, 2,\ldots.
\end{align}
We wish to show that
\[
\lim_{n\to\infty} R_n = \lim_{n\to\infty} \frac{1}{2n} J_n(P, \ldots,
P) = -\log x_0,
\]
where $x_0$ is the unique positive root of the fourth order polynomial
\begin{equation}
\label{ar1 cap2}
P\,x^2 = \frac{(1-x^2)}{(1 + |\alpha| x)^2}.
\end{equation}

Define
\[
\phi(\xi, \zeta_1, \zeta_2) = 
\frac{1}{2} \log (1 + (\sqrt{\zeta_1} + |\alpha| \sqrt{\zeta_2} e^{-\xi})^2),
\qquad \xi, \zeta_1, \zeta_2 \ge 0,
\]
and
\[
\psi(\xi, \zeta) = \phi(\xi, \zeta, \zeta),
\qquad \xi, \zeta \ge 0.
\]
It is easy to check the followings:
\setenumerate{topsep=0em, partopsep=0em}
\begin{enumerate}[label=(\roman*), leftmargin=*]
\item $\phi(\xi, \zeta_1, \zeta_2)$ is increasing and concave in 
$(\zeta_1, \zeta_2)$;
\item for each $\zeta_1, \zeta_2 \ge 0$, $\phi(\xi, \zeta_1, \zeta_2)$
is a decreasing contraction of $\xi$ in the sense that
\[
\phi(\xi_1, \zeta_1, \zeta_2) - \phi(\xi_2, \zeta_1, \zeta_2)
\le \xi_2 - \xi_1
\]
for all $\xi_1$ and $\xi_2$; and consequently,
\item for each $\zeta > 0$, there is a unique solution $\xi^*(\zeta)$
to the equation $\xi = \psi(\xi,\zeta)$ such that $\psi(\xi,\zeta) >
\xi$ for all $\xi < \xi^*(\zeta)$ and $\psi(\xi) < \xi$ for all $\xi >
\xi^*(\zeta)$.
\end{enumerate}
For an arbitrary infinite sequence $\{P_i\}_{i=0}^\infty$ with $P_0 = 0$
and 
\begin{gather}
\label{ar1 power}
\limsup_{n\to\infty}\frac{1}{n} \sum_{i=1}^n P_i \le P,
\end{gather}
we define
\begin{align*}
\xi_0 &= 0,\\
\xi_i &= \phi(\xi_{i-1}, P_i, P_{i-1}), \qquad i = 1,2,\ldots.
\end{align*}
Then we can rewrite the recursion formula \eqref{ar1
recursion1} -- \eqref{ar1 recursion3} as
\begin{align*}
\xi_1 &= \frac{1}{2} J_1(P_1),\\
\xi_i &= \frac{1}{2} \left(
J_i(P_1,\ldots,P_i) - J_{i-1}(P_1,\ldots,P_{i-1})\right),
\qquad i=2,3,\ldots,
\end{align*}
and we have
\begin{align*}
\frac{1}{2n} J_n(P_1,\ldots, P_n) &= \frac{1}{n} \sum_{i=1}^n \xi_i.
\end{align*}

Now we show that
\[
\xi^{**} := \limsup_{n\to\infty} \frac{1}{n} \sum_{i=1}^n \xi_i \le
\xi^*,
\]
where $\xi^* = \xi^*(P)$ is the unique solution to the equation $\xi =
\psi(\xi, P)$.
Indeed, 
\begin{align*}
\frac{1}{n} \sum_{i=1}^n \xi_i
&= \frac{1}{n} \sum_{i=1}^n \phi(\xi_{i-1}, P_i, P_{i-1})\\
&= 
\frac{1}{n} \sum_{i=1}^n \big (
\phi(\xi_{i-1}, P_i, P_{i-1}) - \phi(\xi^{**}, P_i, P_{i-1}) 
+ \phi(\xi^{**}, P_i, P_{i-1}) \big)\\
&\le 
\frac{1}{n} \sum_{i=1}^n \big (
\xi^{**} - \xi_{i-1} + \phi(\xi^{**}, P_i, P_{i-1}) \big)\\
&\le 
\frac{1}{n} \sum_{i=1}^n (
\xi^{**} - \xi_{i-1} ) + \phi\left(\xi^{**}, 
\frac{1}{n}\sum_{i=1}^n P_i, \frac{1}{n}\sum_{i=1}^n P_{i-1}\right),
\end{align*}
where the first inequality follows from the aforementioned property
(ii) and the second inequality follows from the property (i) and
Jensen's inequality.  By taking limits on both sides, we get from
continuity of $\phi(\xi, \zeta_1, \zeta_2)$ in $(\zeta_1, \zeta_2)$
\[
\xi^{**} \le \phi(\xi^{**}, P, P)
= \psi(\xi^{**}, P),
\]
which, from the property (iii), implies that $\xi^{**} \le \xi^*$.  We
can also check that letting $P_i \equiv P$ for all $i = 1,2,\ldots$
attains $\xi^{**} = \xi^*$ from the property (ii) and the principle of
contraction mappings~\cite[Section 14]{KF 1954}.  (See Figure~\ref{ar
figure} below and the detailed analysis in \cite[Section 5]{Wolfowitz
1975}.)  Thus, we conclude that the supremum of $\limsup_{n\to\infty}
{(2n)}^{-1} J_n(P_1,\ldots, P_n)$ over all infinite power sequences
$\{P_i\}$ satisfying the power constraint \eqref{ar1 power} is
achieved by the uniform power allocation.  From simple change of
variable $x_0 = \exp(-\xi^*)$, we can easily verify
\[
\xi^* = -\log x_0
\]
where $0 <x_0 \le 1$ is the unique positive solution to \eqref{ar1
cap2}.
\begin{figure}[ht]
\label{ar figure}
\begin{center}
\input{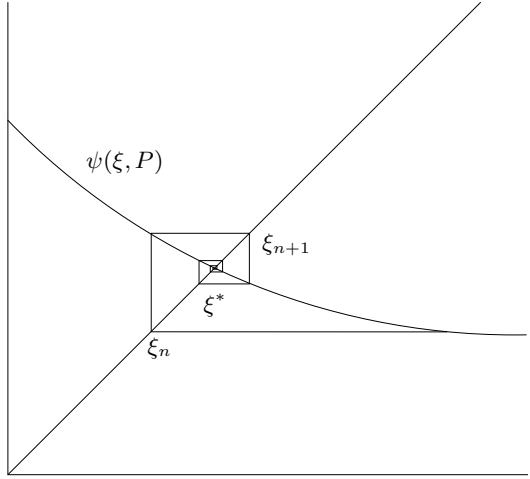}
\caption{Convergence to the unique point $\xi^*$.}
\end{center}
\end{figure}

As in the MA(1) case before, it remains to justify the interchange of
the order of limit and supremum in
\begin{align*}
\lim_{n\to\infty} R_n &=
\lim_{n\to\infty} \max_{P_i: \sum_i P_i \le nP} \frac{1}{2n}
J_n(P_1,\ldots, P_n)\\
&= \sup_{\{P_i\}_{i=1}^\infty}
\limsup_{n\to\infty} \frac{1}{2n} J_n(P_1,\ldots,P_n) \\
&= \lim_{n\to\infty} \frac{1}{2n} J_n(P,\ldots,P) \\
&= \xi^*.
\end{align*}
Obviously we have
\[
\lim_{n\to\infty} \max_{P_i: \sum_i P_i \le nP} \frac{1}{2n}
J_n(P_1,\ldots, P_n) \ge \lim_{n\to\infty} \frac{1}{2n}
J_n(P,\ldots,P).
\]
For the other direction of inequality, fix $n$ and take $(P_1^*,
\ldots, P_{n-1}^*)$ that achieves $R_{n-1}$.  We construct the
infinite sequence $\{P_i\}_{i=1}^\infty$ by concatenating $(P_1^*,
\ldots, P_{n-1}^*, 0)$ repeatedly, that is, $P_{kn + i} = P_i^*,$ $1
\le i \le n-1, k = 0, 1,\ldots,$ and $P_{kn} = 0$ for all $k =
1,\ldots.$ Now we can easily verify that
\[
J_{kn}(P_1,\ldots,P_{kn}) = k J_{n-1}(P_1^*, \ldots, P_{n-1}^*)
= 2k(n-1) R_{n-1}.
\]
(Taking $P_{kn} = 0$ resets the dependence on the past.)  By taking
limits on both sides, we get
\begin{align*}
\lim_{n\to\infty} R_{n-1} &= \lim_{n\to\infty} \frac{n}{n-1}
\frac{1}{2kn} J_{kn} (P_1,\ldots, P_{kn}) \\
&\le \limsup_{n\to\infty} \frac{1}{2n} J_n (P_1,\ldots, P_n)\\
&\le \lim_{n\to\infty} \frac{1}{2n} J_n (P,\ldots, P).
\end{align*}
This completes the proof of the asymptotic optimality of the uniform
power allocation.\qed

\section*{Acknowledgement}
The author is pleased to express his gratitude to Tom Cover for his
invaluable insights and guidance throughout this work.  He thanks
Styrmir Sigurj\'onsson and Erik Ordentlich for enlightening
discussions, and Sina Zahedi for his numerical optimization program,
which was especially useful in the initial phase of this study.  He is
also grateful to anonymous reviewers for their careful reading of the
paper, which resulted in many improvements of the manuscript.

\end{document}